\documentclass[nohyper,12pt,letterpaper]{JHEP3}
\usepackage{epsfig}
\usepackage[latin1]{inputenc}
\usepackage{bbm,amsfonts}
\usepackage{graphicx}
\usepackage{amssymb,amsmath}

\author{Marco S. Bianchi$^\ast$,
  Matias Leoni$^{\ast}$,
  Andrea Mauri$^{\dag,\hash}$,
  Silvia Penati$^\ast$
  and Alberto Santambrogio$^\hash$\\\\
  $^\ast$Dipartimento di Fisica, Universit\`a di Milano--Bicocca and
  INFN, Sezione di Milano--Bicocca, Piazza della Scienza 3, I-20126 Milano, Italy \\\\
  $^\dag$Dipartimento di Fisica dell'Universit\`a degli studi di Milano\\\\
  $^\hash$ INFN, Sezione di Milano, via Celoria 16, I-20133 Milano, Italy
  \qquad\\\\
  E-mail: \email{marco.bianchi@mib.infn.it, matias.leoni@mib.infn.it,
    andrea.mauri@mi.infn.it, silvia.penati@mib.infn.it,
    alberto.santambrogio@mi.infn.it }}

\abstract{We study the correspondence between scattering amplitudes and Wilson loops in three--dimensional Chern--Simons matter theories. In particular, using $\mathcal{N}=2$ superspace formalism, we compute at one loop the whole spectrum of four--point superamplitudes for generic $\mathcal{N} \geq 2$ supersymmetric theories and find a vanishing result for $\mathcal{N}=6$ ABJ(M) and $\mathcal{N}=8$ BLG models.  This restricts the possible range of theories for which Wilson loops/scattering amplitudes duality might work. At two loops, we present the computation of the four-point ABJ scattering amplitude for external chiral superfields. Extending the known result for the ABJM Wilson loop to the ABJ case we find  perfect agreement. We also discuss the dual conformal invariance of our results and the relationship between the Feynman diagram computation and unitarity methods. While for the ABJM theory dual conformally invariant integrals exactly reproduce the amplitude, for the ABJ case this happens only up to a residual constant depending on the parity--violating parameter. Finally we propose a BDS--like exponentiation for the amplitude based on an analogy with the four dimensional $\mathcal{N}=4$ SYM case, and discuss its strong coupling dual counterpart.}

\preprint{October 2011\\ IFUM-982-FT}

\title{SCATTERING IN ABJ THEORIES}


\csname @addtoreset\endcsname{equation}{section}

\def\bseq{\begin{subequation}}  
\def\eseq{\end{subequation}}
\def\bsea{\begin{subeqnarray}}  
\def\esea{\end{subeqnarray}}

\newcommand{\beq}{\begin{equation}}
\newcommand{\bea}{\begin{eqnarray}}
\newcommand{\eea}{\end{eqnarray}}
\newcommand{\eeq}{\end{equation}}

\newcommand {\non}{\nonumber}

\newcommand{\Ab}{\bar{A}}
\newcommand{\Bb}{\bar{B}}
\newcommand{\hb}{\bar{h}}

\renewcommand{\a}{\alpha}
\renewcommand{\b}{\beta}

\renewcommand{\d}{\delta}
\newcommand{\pa}{\partial}
\newcommand{\g}{\gamma}
\newcommand{\G}{\Gamma}

\newcommand{\e}{\epsilon}
\newcommand{\z}{\zeta}

\renewcommand{\l}{\lambda}
\renewcommand{\L}{\Lambda}

\newcommand{\p}{\pi}

\newcommand{\s}{\sigma}

\newcommand{\Db}{\overline{D}}

\newcommand{\thb}{\overline{\theta}}

\renewcommand{\thb}{\overline{\theta}}

\newcommand{\intke}[1]{\int\!\! \frac{d^{3-2\e}{#1}}{(2\pi)^{3-2\e}} ~}

\def\Mb{\kern 2pt\mathchoice
        {
         \vbox{\hrule width10pt height 0.4pt depth 0pt
         \kern 1.2pt\hbox{\kern -2pt$\displaystyle M$}}}
        {
         \vbox{\hrule width10pt height 0.4pt depth 0pt
         \kern 1.2pt\hbox{\kern -2pt$\textstyle M$}}}
        {
\vbox{\hrule width6pt height 0.4pt depth 0pt
         \kern 1.0pt\hbox{\kern -2pt$\scriptstyle M$}}}
        {
         \vbox{\hrule width5pt height 0.4pt depth 0pt
         \kern 0.8pt\hbox{\kern -2pt$\scriptscriptstyle M$}}}}

\def\Sb{\kern 2pt\mathchoice
        {
         \vbox{\hrule width6pt height 0.4pt depth 0pt
         \kern 1.2pt\hbox{\kern -2pt$\displaystyle S$}}}
        {
         \vbox{\hrule width6pt height 0.4pt depth 0pt
         \kern 1.2pt\hbox{\kern -2pt$\textstyle S$}}}
        {
         \vbox{\hrule width3.5pt height 0.4pt depth 0pt
         \kern 1.0pt\hbox{\kern -2pt$\scriptstyle S$}}}
        {
         \vbox{\hrule width3pt height 0.4pt depth 0pt
         \kern 0.8pt\hbox{\kern -2pt$\scriptscriptstyle S$}}}}

\def\Rb{\kern 2pt\mathchoice
        {
         \vbox{\hrule width5.5pt height 0.4pt depth 0pt
         \kern 1.2pt\hbox{\kern -2.5pt$\displaystyle R$}}}
        {
         \vbox{\hrule width5.5pt height 0.4pt depth 0pt
         \kern 1.2pt\hbox{\kern -2.5pt$\textstyle R$}}}
        {
         \vbox{\hrule width3.5pt height 0.4pt depth 0pt
         \kern 1.0pt\hbox{\kern -2.2pt$\scriptstyle R$}}}
        {
         \vbox{\hrule width3pt height 0.4pt depth 0pt
         \kern 0.8pt\hbox{\kern -2.2pt$\scriptscriptstyle R$}}}}

  \def\pp{{\mathchoice
      {
          \kern 1pt%
          \raise 1pt
          \vbox{\hrule width5pt height0.4pt depth0pt
            \kern -2pt
            \hbox{\kern 2.3pt
              \vrule width0.4pt height6pt depth0pt
              }
            \kern -2pt
            \hrule width5pt height0.4pt depth0pt}%
            \kern 1pt
       }
        {
          \kern 1pt%
          \raise 1pt
          \vbox{\hrule width4.3pt height0.4pt depth0pt
            \kern -1.8pt
            \hbox{\kern 1.95pt
              \vrule width0.4pt height5.4pt depth0pt
              }
            \kern -1.8pt
            \hrule width4.3pt height0.4pt depth0pt}%
            \kern 1pt
        }
        {
          \kern 0.5pt%
          \raise 1pt
          \vbox{\hrule width4.0pt height0.3pt depth0pt
            \kern -1.9pt  
            \hbox{\kern 1.85pt
              \vrule width0.3pt height5.7pt depth0pt
              }
            \kern -1.9pt
            \hrule width4.0pt height0.3pt depth0pt}%
            \kern 0.5pt
        }
        {
          \kern 0.5pt%
          \raise 1pt
          \vbox{\hrule width3.6pt height0.3pt depth0pt
            \kern -1.5pt
            \hbox{\kern 1.65pt
              \vrule width0.3pt height4.5pt depth0pt
              }
            \kern -1.5pt
            \hrule width3.6pt height0.3pt depth0pt}%
            \kern 0.5pt
        }
    }}

  \def\mm{{\mathchoice
               {
                 \kern 1pt
           \raise 1pt    \vbox{\hrule width5pt height0.4pt depth0pt
                  \kern 2pt
                  \hrule width5pt height0.4pt depth0pt}
                 \kern 1pt}
               {
                \kern 1pt
           \raise 1pt \vbox{\hrule width4.3pt height0.4pt depth0pt
                  \kern 1.8pt
                  \hrule width4.3pt height0.4pt depth0pt}
                 \kern 1pt}
               {
                \kern 0.5pt
           \raise 1pt
                \vbox{\hrule width4.0pt height0.3pt depth0pt
                  \kern 1.9pt
                  \hrule width4.0pt height0.3pt depth0pt}
                \kern 1pt}
               {
               \kern 0.5pt
         \raise 1pt  \vbox{\hrule width3.6pt height0.3pt depth0pt
                  \kern 1.5pt
                  \hrule width3.6pt height0.3pt depth0pt}
               \kern 0.5pt}
               }}

\def\pd{{\kern0.5pt
           + \kern-5.05pt \raise5.8pt\hbox{$\textstyle.$}\kern
0.5pt}}

\def\pmd{{\kern0.5pt
          \pm \kern-5.05pt
\raise6.3pt\hbox{$\textstyle.$}\kern1.5pt}}

\def\md{{\mathchoice
   {
      {{\kern 1pt - \kern-6.2pt \raise5pt\hbox{$\textstyle.$}\kern
1pt}}}
    {
      {{\kern 1pt - \kern-6.2pt \raise5pt\hbox{$\textstyle.$}\kern
1pt}}}
    {
      {\kern0.5pt - \kern-5.05pt
\raise3.4pt\hbox{$\textstyle.$}\kern0.5pt}}
    {
      {\kern0.5pt - \kern-5.05pt
\raise3.4pt\hbox{$\textstyle.$}\kern0.5pt}}}}

\def\beq{\begin{equation}}
\def\eeq{\end{equation}}
\def\bea{\begin{eqnarray}}
\def\eea{\end{eqnarray}}
\def\Tr{\textstyle{Tr}}
\def\tr{\textstyle{tr}}
\def\a{\alpha}
\def\b{\beta}
\def\g{\gamma}

\def\d{\delta}
\def\e{\epsilon}
\def\z{\zeta}
\def\th{\theta}
\def\l{\lambda}
\def\G{\Gamma}

\def\L{\Lambda}

\begin{document}

\section{Introduction}

Recently, new interest has been devoted to the study of the S--matrix for the non--trivial sector of three dimensional Chern--Simons--matter theories which allow for a string theory dual description. These are the well--known ${\cal N}=6$ superconformal ABJM model \cite{ABJM} for $U(N)_K \times U(N)_{-K}$ gauge group and the more general ABJ model \cite{ABJ} for  $U(M)_K \times U(N)_{-K}$, where $K$ is the Chern--Simons level. In the large $M,N$ limit their strong dual description is given in terms of M--theory on ${\rm AdS}_4 \times S^7/Z_K$ background and, for $K \ll N \ll K^5$, by a type IIA string theory on ${\rm AdS}_4 \times {\rm CP}_3$.

The main motivation is to understand whether these theories, even if distinguished in nature being them non--maximally supersymmetric, share
fundamental properties of the four dimensional ${\cal N}=4$ SYM theory, like integrability \cite{integrability}, Yangian symmetry \cite{DHP} of the planar physical sector and (scattering amplitudes)/(Wilson loop (WL))/(correlation functions) dualities \cite{Drummond:2007au}--\cite{Belitsky:2011zm}, \cite{AEKMS}--\cite{Adamo:2011dq}.
While going deep into the nature of three dimensional theories, the investigation of these properties should
help  to understand their actual origin and the role of the AdS/CFT in their determination.

For the ABJM model, preliminary results can be already found in literature, concerning integrability \cite{MZ}--\cite{LMMSSST}, the related Yangian symmetry \cite{BLM, Lee} and dualities.

At classical level, scattering amplitudes have been shown to be invariant under dual superconformal symmetry \cite{HL, GHKLL} whose generators are the level--one generators of a Yangian symmetry \cite{BLM}.
At strong coupling this symmetry should rely on self--duality properties of type IIA string on  ${\rm AdS}_4 \times {\rm CP}_3$ under a suitable combination of bosonic and fermionic T--dualities \cite{AM, BM}, even if the situation is complicated by the emergence of singularities in the fermionic T--transformations \cite{ADO}--\cite{DO}.  Considerable progress in this direction has been recently done in \cite{BOY}.

At quantum level, first  evidence of the existence of dualities and the persistence of dual conformal invariance comes from recent findings on scattering amplitudes, light--like WL and correlators of BPS operators at one and two loops.

At one loop both the  four--point amplitude \cite{ABM} and the light--like four--polygon WL \cite{HPW, BLMPRS} vanish. For ${\cal N}=2,3,6,8$ Chern--Simons--matter theories, correlation functions of $2n$ BPS operators have been computed \cite{BLMPRS}
\footnote{The result of \cite{BLMPRS} holds also for the ${\cal N}=8$ BLG theory \cite{BL,G} described by 
$SU(2)_K \times SU(2)_{-K}$ Chern--Simons--matter theory in the large $K$ limit. Enhancement to maximal supersymmetry could also be obtained for Chern--Simons levels $K=1,2$. However, these values are out of the perturbative regime and will not be considered.}.
It has been proved that the one--loop result divided by the corresponding tree level expression coincides with the one--loop light--like $2n$--polygon WL \cite{HPW}. The identification is at the level of the integrands, independently of the fact that both of them eventually vanish.

Less trivial evidence for a (scattering amplitude)/WL duality arises at two loops where these quantities are not supposed to vanish.
Very recently, it has been proved that for the ABJM theory, at this order the four--point scattering amplitude \cite{CH, BLMPS1} divided by its tree--level counterpart coincides with the second order expansion of a light--like four--polygon Wilson loop \cite{HPW}.

In this paper we give details of the calculation for the four--point scattering amplitude up to two loops and extend the result to the more general ABJ theory.

Using ${\cal N}=2$ superspace description and a direct Feynman diagram approach, at one loop we compute the whole spectrum of four--point superamplitudes for ${\cal N}= 2,3,6,8$ Chern--Simons matter conformal field theories with $U(M) \times U(N)$ gauge group. The result is generically different from zero, except for the ${\cal N}=6$ ABJ(M) and ${\cal N}=8$ BLG cases where they all vanish. Suitably generalizing the definition of Wilson loop to the ABJ theory, we easily argue that it also vanishes at one loop. Therefore, we conclude that a (scattering amplitude)/WL duality may work only for the ${\cal N} \geq 6$ case, independently of the fact that conformal symmetry is present in all the theories we analyze.

Focusing on ABJ models, still using a direct Feynman diagram approach, we evaluate the two--loop planar scattering superamplitude of four chiral superfields, two of them in the bifundamental and two in the antibifundamental representation of the  $U(M) \times U(N)$ gauge group. The result for the ratio $\mathcal{M}^{(2)}_4 \equiv \mathcal{A}_4^{(2 \, loops)}/\mathcal{A}^{tree}_4$ is
\begin{equation}
\label{2loop}
\mathcal{M}^{(2)}_4 =  \lambda \hat{\lambda} 
\, \left[ -\frac{( s/\mu'^2)^{-2\epsilon}}{(2\, \epsilon)^2}-\frac{(t/\mu'^2)^{-2\epsilon}}{(2\, \epsilon)^2}+\frac12\,\ln^2 \left(\frac{s}{t}\right)+ C_{\cal A}(M,N)  +
\mathcal{O}(\epsilon)\right]
\end{equation}
where $\lambda=M/K$, $\hat{\lambda}=N/K$, $\mu'$ is the mass scale and $C_{\cal A}(M,N)$ is a constant depending on the ranks of the groups. For $M=N$ we are back to the result for the ABJM theory \cite{CH, BLMPS1}.

This result has a number of remarkable properties.

First of all,  up to an additive, scheme--dependent constant, this expression matches exactly the second order expansion of a ABJ light--like four--polygon Wilson loop, once the IR regularization is formally identified with the UV one and the particle momenta are expressed in terms of dual coordinates ($s= x_{13}^2$ and $t = x_{24}^2$). Therefore, at least for the four--point amplitude, there is evidence that the following identity
\beq
\ln{ {\cal M}_4} = \ln \langle W_4 \rangle + {\rm const.}
\eeq	
should hold order by order in the perturbative expansion of the two objects.

The result (\ref{2loop}) can be identified with the first order expansion of a BDS--like ansatz for the ABJ(M) model
\begin{equation}
\label{BDSlike}
\mathcal{M}_4   = e^{Div + \frac{f_{CS}(\l, \hat{\l})}{8}\left(\ln^2\left(\frac{s}{t}\right)+ \frac{4 \pi^2}{3}\right)  + C(\lambda, \hat{\l}) }
\end{equation}
where $C(\l, \hat{\l})$ is a scheme--dependent constant. This ansatz  is exactly the BDS ansatz \cite{BDS, Bern:2008ap} for ${\cal N}=4$ SYM where the four dimensional scaling function has been substituted by the three dimensional one, $f_{CS}(\lambda, \hat{\l})$, which is an obvious generalization of the ABJM scaling function obtained from the conjectured asymptotic Bethe equations \cite{GV}.

In the ${\cal N}=4$ SYM case, the BDS exponentiation of scattering amplitudes holds also at strong coupling where, according to the Alday-Maldacena prescription \cite{AM}, the amplitudes are given by the exponential of a minimal--area surface in the ${\rm AdS}_5$ dual background ending on a light--like polygon, whose edges are determined by the particle momenta.  In particular, since this prescription is equivalent to computing a light--like WL in AdS, it supports the amplitude/WL duality at strong coupling, 
in agreement with the findings at weak coupling.  

The natural question which arises is whether a similar prescription at strong coupling can be formulated for the ABJ(M) models.  Motivated by the evidence in favor of the amplitudes/WL duality at weak coupling and BDS exponentiation, we expect it to be the case.  
In fact, focusing on the ABJM theory in the intermediate regime $K \ll N \ll K^5$, we discuss the generalization of the Alday-Maldacena prescription to ${\rm AdS}_4 \times {\rm C}P_3$. We find that, apart from a rigorous prescription for the regularization procedure in AdS that we have not formulated properly, the five dimensional solution can be adapted to the four dimensional case and the output is an expression for the four--point amplitude given by eq. (\ref{BDSlike}) where the scaling function assumes its leading value at strong coupling, $f_{CS}(\l) \sim \sqrt{2\l}$. 

For $M=N$, our result (\ref{2loop}) coincides with the one in \cite{CH} obtained by making the ansatz that dual conformal invariance should hold also at loop level. Therefore our calculation supports that ansatz and provides a direct proof of the assumption that dual conformal invariance should be the correct symmetry principle to select the scalar master integrals contributing to the on--shell sector of the theory.

In fact, for the ABJM case, following \cite{CH} we can rewrite the result (\ref{2loop}) as a linear combination of scalar momentum integrals which are dual to three dimensional true conformally invariant integrals, well defined off--shell. As a consequence, the four--point amplitude satisfies anomalous Ward identities associated to dual conformal transformations \cite{Drummond:2007au}, as dual conformal invariance is broken in the on--shell limit by the appearance of IR divergences which require introducing a mass regulator. 

For the ABJ model the situation is slightly complicated by the appearance of a non--trivial dependence 
on the parity--violating parameter $\s = (M-N)/\sqrt{MN}$ in the mass--scale and in the constant $C_{\cal A}$. 
In fact, for $M \neq N$ the two--loop ratio (\ref{2loop}) can be still written as a linear combination of dual invariant integrals only up to an additive constant proportional to $\s^2$. Therefore, for the ABJ theory the dual conformal invariance principle combined with the unitarity cuts method is not sufficient to uniquely fix the amplitude already in the case of four external particles. 

Having computed the two--loop amplitude by a genuine perturbative approach without any {\em a priori} ansatz on its form, we can investigate whether dual conformal properties can be detected even at the level of Feynman diagrams. We have then studied the two loop diagrams entering our calculation, out of the mass--shell and in three dimensions.  Since in three dimensions dual conformal symmetry rules out bubbles, it is immediate to realize that, being some of our diagrams built by bubbles, it cannot work at the level of the integrand on every single diagram. A less stringent scenario could still allow for the possibility to see dual conformal invariance realized at the level of the integrals and after summation of all the contributions. We have made many numerical checks but the output is always negative: dual conformal invariance is definitively broken at the level of Feynman diagrams. However, this is not in contrast with what claimed before, since in three dimensions and in dimensional regularization the integrals do not have in general a smooth limit on the mass--shell.

The organization of the paper is the following.  In Section 2 we briefly review the ABJ(M) theory in ${\cal N}=2$ superspace formalism. In Section 3 we define light--like Wilson loops for the ABJ theory by suitably generalizing the ones for ABJM, and compute  the two--loop expansion for the four--cusp case.  Section 4 contains our main results. There, we give details of the one--loop calculation of the four--point scattering amplitudes in ${\cal N}\geq 2$ theories and of the two--loop ${\cal N}=6$ ABJ amplitude for four chiral superfields. A detailed discussion of our results is given in Section 5 where we analyze the amplitude/WL duality and the related dual conformal invariance of the two--loop amplitude, we make a conjecture on its exponentiation  and investigate the consequences on its strong coupling dual description. Perspectives and open questions are also highlighted. Three Appendices fix the notations and provide technical tools for carrying on the calculations.

\section{Generalities on ABJ theories}

In three dimensions, we consider ${\cal N}=2$ supersymmetric Chern--Simons
theories for $U(M) \times U(N)$ gauge group, generically coupled to chiral matter.
In ${\cal N}=2$ superspace the field content is organized into two vector
multiplets $(V,\hat{V})$ in the adjoint representation of
the gauge groups $U(M)$ and $U(N)$ respectively, and
four chiral multiplets $A^i$ and $B_i$, $i=1,2$ , with $A^i$ in
the $(M,\bar{N})$ and $B_i$ in the $(\bar{M},N)$ bifundamental
representations.

We consider the action (for superspace conventions see Appendix A)
\begin{equation} {\cal S} = {\cal S}_{\mathrm{CS}} + {\cal
    S}_{\mathrm{mat}}
  \label{eqn:action}
\end{equation}
with
\begin{eqnarray}
  \label{action}
  && {\cal S}_{\mathrm{CS}}
  =  \int d^3x\,d^4\theta \int_0^1 dt\: \Big\{ \frac{K_1}{4\pi}  {\Tr \Big[
  V \Db^\a \left( e^{-t V} D_\a e^{t V} \right) \Big]}
  + \frac{K_2}{4\pi}  \Tr \Big[ \hat{V} \Db^\a \left( e^{-t \hat{V}} D_\a
    e^{t \hat{V}} \right) \Big]   \Big\}
  \non \\
  \non \\
  && {\cal S}_{\mathrm{mat}} = \int d^3x\,d^4\theta\: \Tr \left( \bar{A}_i
    e^V A^i e^{- \hat{V}} + \bar{B}^i e^{\hat V} B_i
    e^{-V} \right)
  \non \\
  &~& ~ ~~~\qquad ~+\int d^3x\,d^2\theta\:
  \Tr  \left[  h_1 (A^1 B_1 A^2 B_2) + h_2 (A^2 B_1 A^1 B_2) \right]  + \, h.c.
\end{eqnarray}
Here $h_i$ are two generic complex couplings, while $K_1, \, K_2$ are two independent integers as required
by gauge invariance of the effective action.  In the perturbative regime we take $K_1, K_2 \gg M,N$.

For special values of the $h_i$'s we can have enhancement of global
symmetries and/or R--symmetry with consequent enhancement of
supersymmetry \cite{Klebanov}.

For $K_1= - K_2 \equiv K$ and  $h_1 = - h_2$, the global symmetry
becomes $U(1)_R \times SU(2)_A \times SU(2)_B$ and gets enhanced to
$SU(4)_R$ for \cite{ABJM, Klebanov}
\beq
\label{ABJ}
h_1 = - h_2 = 4\pi/K
\eeq
For this particular value of the couplings we recover the ${\cal N}=6$ superconformal ABJ
theory \cite{ABJ} and for $N=M$ the ABJM theory \cite{ABJM}.

Non--trivial fixed points can be found also for $K_1 \neq -K_2$ \cite{BPS1,BPS2}.
In the large $M,N$ limit, choosing
\beq
\label{SU(2)}
h_1 = -h_2 = 4\pi \sqrt{\frac{1}{K_1^2} +  \frac{1}{K_2^2} +\frac{1}{K_1 K_2}}
\eeq
we find ${\cal N}=2$ superCFT's with $SU(2)_A \times SU(2)_B$ global symmetry.
These theories are dual to string backgrounds where a Romans mass has been turned on \cite{GT}.

More generally,  we can take $h_1 \neq -h_2$ satisfying in the large $M,N$ limit
\beq
\label{U(1)}
|h_1|^2 + |h_2|^2 = 32 \pi^2 \left( \frac{1}{K_1^2} +  \frac{1}{K_2^2} +\frac{1}{K_1 K_2} \right)
\eeq
This corresponds to a  set of ${\cal N}=2$ superCFT's with $U(1)_A \times U(1)_B$ global symmetry.

\vskip 10pt
The quantization of these theories can be easily carried on in superspace
after performing gauge fixing (for details, see for instance
\cite{BPS1, BPS2, AndreaMati}). In dimensional regularization, $d=3-2\e$,  and using Landau gauge, this
leads to gauge propagators (in configuration and momentum spaces)
\begin{eqnarray}
 \langle V^A(1) \, V^B(2) \rangle
&=&   \frac{1}{K_1} \, \frac{\G(1/2 -\e)}{\pi^{1/2-\e}} \, \Db^\a D_\a \, \frac{\delta^4(\th_1-\th_2)}{|x_1 - x_2|^{1-2\e}} \; \delta^{AB}
\non \\
&\longrightarrow &
  \frac{4\pi}{K_1} \, \frac{1}{p^2}  \; \Db^\a D_\a \, \delta^4(\th_1-\th_2) \; \delta^{AB}
  \nonumber \\
  \non \\
  \langle \hat V^A(1) \, \hat V^B(2) \rangle &=&
  \frac{1}{K_2} \, \frac{\G(1/2 -\e)}{\pi^{1/2-\e}} \, \Db^\a D_\a  \, \frac{\delta^4(\th_1-\th_2)}{|x_1 - x_2|^{1-2\e}} \; \delta^{AB}
  \non \\
&\longrightarrow &
  \frac{4\pi}{K_2} \,  \frac{1}{p^2}   \; \Db^\a D_\a  \, \delta^4(\th_1-\th_2) \; \delta^{AB}
  \label{gaugeprop}
\end{eqnarray}
Analogously, the scalar propagators are
\begin{eqnarray}
 \langle \bar{A}^{\hat a}_{\ a}(1) \, A^b_{\ \hat b}(2) \rangle &  = &
\frac{\G(1/2 -\e)}{4 \pi^{3/2-\e}} \, \frac{\delta^4(\th_1 - \th_2)}{|x_1 - x_2|^{1-2\e}} \; \delta^{\hat a}_{\ \hat
    b} \, \delta^{\ b}_{a}
\non \\
& \longrightarrow &
 \frac{1}{p^2}   \; \delta^4(\th_1 - \th_2) \, \, \delta^{\hat a}_{\ \hat
    b} \, \delta^{\ b}_{a}
    \nonumber \\
    \non \\
 \langle \bar{B}^a_{\ \hat a}(1) \, B^{\hat b}_{\ b}(2) \rangle &=&
\frac{\G(1/2 -\e)}{4 \pi^{3/2-\e}} \,  \frac{\delta^4(\th_1 - \th_2)}{|x_1 - x_2|^{1-2\e}}  \; \delta^a_{\ b} \, \delta^{\ \hat b}_{\hat a}
\non \\
&\longrightarrow&
 \frac{1}{p^2}  \; \delta^4(\th_1 - \th_2) \, \,  \delta^a_{\ b} \, \delta^{\ \hat b}_{\hat a}
  \label{scalarprop}
\end{eqnarray}
where $a,b$ and $\hat{a}, \hat{b}$ are indices of the (anti)fundamental representation of $U(M)$ and $U(N)$, respectively.

The vertices needed for loop calculations can be easily read from the action (\ref{action}) when expanded in powers of $(V, \hat{V})$ up to the desired order.

The renormalization of the effective action $e^\G \equiv \int e^S$, has been studied up to two loops
in Refs. \cite{BPS1, BPS2} using dimensional regularization with dimensional reduction (see also \cite{Akerblom:2009gx}).
At one loop, there are only finite corrections to the gauge quadratic action which in the large $M,N$ limit read
\bea
\label{1loopprop}
&&    \G^{(1)}_{gauge} =  G[1,1] \, \left( N - \frac{M}{4} \right) \int  \frac{d^3 p}{(2\pi)^3} \,d^4\th \;
{\Tr \Big( V(p)} \frac{\Db^\a D^2 \Db_\a}{|p|^{1+2\epsilon}} V(-p) \Big)
\non \\
&&  \hat{\G}^{(1)}_{gauge} =  G[1,1] \, \left( M -\frac{N}{4} \right) \int   \frac{d^3 p}{(2\pi)^3} \
d^4\th \;  {\Tr \Big( \hat{V}(p)} \frac{\Db^\a D^2 \Db_\a}{|p|^{1+2\epsilon}} \hat{V}(-p) \Big)
\eea
where the $G$ function is defined in (\ref{G}).
At two loops, UV divergences appear which lead to non--trivial beta--functions. The condition for them to vanish determines the conformal fixed points listed in Eqs. (\ref{ABJ}, \ref{SU(2)}, \ref{U(1)}).

\section{Light--like Wilson loops for ABJ theories}

The four--cusp Wilson loop for ABJM theory with $U(N) \times U(N)$ gauge group proposed in \cite{DPY, HPW} is
\bea\label{WL}
\langle W_4 \rangle_{\text{ABJM}} = \frac{1}{2N}\,\left\{ \Tr e^{\, i \int_{\g} A_{\mu}\, d\, z^{\mu}} + \Tr  e^{\, i \int_{\g} \hat{A}_{\mu}\, d\, z^{\mu}} \right\}
\eea
where $\g$ is a light--like four--polygon closed path.

The perturbative evaluation has revealed that this expression vanishes at one loop, while at two loops it gets non--trivial contributions both from the gauge and the matter sectors \cite{HPW}. In the large $N$ limit
one can write
\beq
\langle W_4 \rangle^{(2)}_{\text{ABJM}} = N^2 \, \left[ \langle W_4 \rangle^{(2)}_{CS} + \langle W_4 \rangle^{(2)}_{matter} \right]
\eeq
where, up to ${\cal O}(\e)$ terms,
\bea
\label{CS}
\langle W_4 \rangle^{(2)}_{CS} =
- \frac{1}{K^2}\,  \left[ \frac{1}{2} \ln{2} \, \frac{(x_{13}^2 \, \pi e^{\gamma_E} \mu^2  )^{2\epsilon}+
  (x_{24}^2 \, \pi e^{\gamma_E} \mu^2 )^{2\epsilon}}{\epsilon} +  \frac{1}{4} (a_6 - 8 \ln{2} - \pi^2) \right]
\eea
$a_6$ being a constant determined numerically (see Ref. \cite{HPW}), and
\bea
\label{matter}
\langle W_4 \rangle^{(2)}_{matter} =
- \frac{1}{K^2}\, \left[  \frac{(x_{13}^2 \, 4 \pi e^{\gamma_E}\mu^2 \, )^{2\epsilon}}{(2\epsilon)^2} +
\frac{(x_{24}^2 \, 4 \pi e^{\gamma_E}\mu^2 \,)^{2\epsilon}}{(2\epsilon)^2}  - \frac12\,  \ln^2\left(\frac{x_{13}^2}{x_{24}^2}\right) - \frac{\pi^2}{4}  \right]
\eea
Here $\g_E$ is the Euler constant.
Summing the two contributions, one obtains
\begin{align}
\label{WLABJM}
\langle W_4 \rangle^{(2)}_{\text{ABJM}} = \l^2 \left[   -\frac{(x_{13}^2 \, {\mu}_{WL}^2 )^{2\epsilon}}{(2\epsilon)^2} -
\frac{(x_{24}^2 \, {\mu}_{WL}^2 )^{2\epsilon}}{(2\epsilon)^2}  + \frac12 \,  \ln^2\left(\frac{x_{13}^2}{x_{24}^2}\right)   + C   \right]
\end{align}
where $\l \equiv N/K$, ${\mu}_{WL}^2= 8 \pi e^{\gamma_E}\mu^2 $ and
\beq
C= 3 \z_2 + 2 \ln{2} + 5 \ln^2{2} -\frac{a_6}{4}
\eeq

We generalize the definition (\ref{WL}) to the larger class of ABJ models, taking into account the possibility for the two gauge groups to have different ranks.

It is convenient to introduce two 't Hooft couplings $\lambda=M/K$ and $\hat\lambda=N/K$. It follows that  the perturbative parameter is $\bar\lambda = \sqrt{\lambda\hat\lambda}$, while
\begin{equation}
\label{sigma}
\sigma=\frac{\lambda-\hat\lambda}{\bar\lambda},
\end{equation}
measures the deviation from the ABJM theory.

For gauge group $U(M)\times U(N)$ we consider
\bea\label{WL1}
\langle W_4 \rangle^{(2)}_{\text{ABJ}} = \frac{1}{2M} \,{\Tr_{\text{\scriptsize $U(M)$}} }e^{\, i \int_\g  A_{\mu}\, d\, z^{\mu}} + \frac{1}{2N}\, {\Tr_{\text{\scriptsize $U(N)$}}} e^{\, i \int_\g  \hat{A}_{\mu}\, d\, z^{\mu}}
\eea
Following Ref. \cite{HPW}, it is immediate to realize that this expression is still vanishing at one loop, as
this result is independent of the choice of the gauge group.

At two loops, the contributing diagrams are the same as in the ABJM case, but with different color coefficients.  Generalizing the calculation of \cite{HPW}, in the large $M,N$ limit we obtain
\bea\label{WL1result}
\langle W_4 \rangle^{(2)}_{\text{ABJ}} =  \frac12\, (M^2 +  N^2)\, \langle W_4 \rangle^{(2)}_{CS} + MN\, \langle W_4 \rangle^{(2)}_{matter}
\eea
where $\langle W_4 \rangle^{(2)}_{CS}$ and $\langle W_4 \rangle^{(2)}_{matter}$ are still given in eqs.
(\ref{CS}) and (\ref{matter}), respectively.

Inserting their explicit expressions and rescaling the regularization parameter as
\bea
\label{rescaling1}
\mu'^{\, 2}_{WL} =  2^{3+\sigma^2/2} \, \pi \,  e^{\g_E}\, \mu^2
\eea
up to terms of order $\e$, the final answer reads
\bea
\label{resultWL1}
\langle W_4 \rangle^{(2)}_{\text{ABJ}} = \bar\lambda^2 \Big\{
 - \frac{\left( x_{13}^2\, \mu'^{\, 2}_{WL} \right)^{2 \epsilon }}{(2 \epsilon)^2} - \frac{\left( x_{24}^2\, 
 \mu'^{\, 2}_{WL} \right)^{2 \epsilon }}{(2 \epsilon)^2}
 + \frac12\, \ln^2\left(\frac{x_{13}^2}{x_{24}^2}\right) + C'  \Big\}
\non \\
\eea
where
\bea
\label{const1}
C' &=&  \frac34 \, (\sigma^2+4) \, \z_2  + (\sigma^2+2) \ln{2}
 - \frac{\sigma^2+2}{8}\, a_6
+ \frac{(\sigma^2+2)(\sigma^2+10)}{4}\, \ln^2{2}
\eea
It is straightforward to check that taking $\sigma\to 0$, the above expression reduces to the result (\ref{WLABJM}) for the Wilson loop in ABJM, rescaling (\ref{rescaling1}) and constant (\ref{const1}) included.

It is interesting to note that another possible candidate for the Wilson loop in ABJ theory is
\bea\label{WL2}
\langle W_4 \rangle_{\text{ABJ}} = \frac{1}{M+N}\,\left\{ \Tr_{U(M)} e^{\, i \int_{\g} A_{\mu}\, d\, z^{\mu}} + \Tr_{U(N)} e^{\, i \int_{\g} \hat{A}_{\mu}\, d\, z^{\mu}} \right\}
\eea
This naturally arises when writing the gauge field as a $(M+N)\times (M+N)$ square matrix, ${\cal A} = \text{diag} (A,\hat{A})$.

It is easy to realize that this other expression is still one--loop vanishing. At two loops, it is given by a slightly different  combination of the expressions (\ref{CS}, \ref{matter})
\bea
\label{resultWL2}
\langle W_4 \rangle^{(2)}_{\text{ABJ}} &=& \frac{1}{M+N}\,\left\{ (M^3 +  N^3)\, \langle W_4 \rangle^{(2)}_{CS} + (M^2\,N + N^2\, M)\, \langle W_4 \rangle^{(2)}_{matter} \right\}\non\\
&=& (M^2 -MN + N^2)\, \langle W_4 \rangle^{(2)}_{CS} + MN\, \langle W_4 \rangle^{(2)}_{matter}
\eea
However, provided that we define
 \beq
\label{rescaling2}
{\mu}''^{\, 2}_{WL}= 2^{3+\sigma^2} \, \pi \, e^{\g_E}\,  \mu^2
\eeq
the calculation leads  exactly to the same result as before
\bea\label{WL2result}
\langle W_4 \rangle^{(2)}_{\text{ABJ}} = \bar\lambda^2 \Big\{
 - \frac{\left( x_{13}^2\, \mu''^{\, 2}_{WL}\right)^{2 \epsilon }}{(2 \epsilon)^2} - \frac{\left(x_{24}^2\, 
 \mu''^{\, 2}_{WL}\right)^{2 \epsilon }}{(2 \epsilon)^2} + \frac12 \ln^2\left(\frac{x_{13}^2}{x_{24}^2}\right)  + C''  \Big\}
\non \\
\eea
where now
\bea
\label{const2}
C'' &=&  \frac32 \, (\sigma^2+2) \, \z_2  + 2 (\sigma^2+1) \ln{2}
 - \frac{\sigma^2+1}{4}\, a_6
+ (\sigma^2+1)(\sigma^2+5)\, \ln^2{2}
\eea
Again, taking $\sigma\to 0$ we are back to the ABJM result (\ref{WLABJM}).

Up to two loops, the two definitions (\ref{WL1}) and (\ref{WL2}) for the Wilson loop in ABJ theory differ only by the choice of the mass scale and the scheme--dependent $C$ constants. At this stage we do not have any tool to discriminate between the two.

\section{Four--point scattering amplitudes}

In  Chern--Simons matter theories the only non--trivial scattering amplitudes are scalar matter amplitudes, as the vector fields are not propagating.  In ${\cal N}=2$ superspace language this means having $A$, $B$ and their complex conjugates  as external superfields. Given the structure of the vertices, it is straightforward to
see that only amplitudes with an even number of external legs are non--vanishing. This is consistent
with the requirement for the amplitudes to be Lorentz and dilatation invariant \cite{HL}.

Each external scalar particle carries an on--shell momentum $p_{\a\b}$ ($p^2 =0$), a $SU(2)$ index
and color indices corresponding to the two gauge groups. We classify as
particles the ones carrying $(M, \bar{N})$ indices and antiparticles the ones carrying $(\bar{M}, N)$
indices. Therefore, $(A^i, \bar{B}^j)$ are {\em particles}, whereas  $(B_i, \bar{A}_j)$ are {\em antiparticles}.

We are interested in the simplest non--trivial amplitudes, that is four--point amplitudes. These are chiral
superamplitudes $(A^i B_j A^k B_l)$ and non--chiral superamplitudes  $(A^i \bar{A}_j A^k \bar{A}_l)$,
$(B_i \bar{B}^j B_k \bar{B}^l)$, $(A^i \bar{A}_j \bar{B}^k B_l)$ plus possible permutations.
While for the ABJ(M) theories they can all be obtained from $(A B A B)$ by $SU(4)$ R--symmetry transformations, for more general ${\cal N}=2$ models they are independent objects and need be computed separately.

The color indices can be stripped out, as we can write
\beq
{\cal A}_4 \left( X^{a_1}_{\, \bar{a}_1}\,  Y^{\bar{b}_2}_{\, b_2} \, Z^{a_3}_{\, \bar{a}_3}
W^{\bar{b}_4}_{\, b_4} \right) =
\sum_{\s}
{\cal A}_4(\s(1),   \cdots , \s(4) ) \; \d^{a_{\s(1)}}_{b_{\s(2)}} \, \d^{\bar{b}_{\s(2)}}_{\bar{a}_{\s(3)}}
\, \d^{a_{\s(3)}}_{b_{\s(4)}} \d^{\bar{b}_{\s(4)}}_{\bar{a}_{\s(1)}}
\eeq
where $(X,Z)$ stay generically for $A$ or $\bar{B}$, $(Y,W)$ for $B$ or $\bar{A}$ and the sum is over exchanges of even and odd sites between themselves.
We can then restrict to color--ordered amplitudes ${\cal A}_4(\s(1),   \cdots , \s(4) )$
with a fixed order of the external momenta.

We compute amplitudes perturbatively, by a direct superspace Feynman diagram approach. Precisely,
for four--point amplitudes, we evaluate the effective action quartic in the scalar matter superfields. Since in a scattering process the external fields are on--shell, it is sufficient to evaluate the {\em on--shell} effective action.  This amounts to require the external superfields to satisfy the equations of motion (EOM)
\footnote{The actual EOM as derived from the action (\ref{action}) would be $D^2 A^1 = - \bar{h}_1 \Bb^2 \Ab_2
\Bb^1 - \bar{h}_2\Bb^1 \Ab_2 \Bb^2 $, $D^2 A^2 =  - \bar{h}_1 \Bb^1 \Ab_1 \Bb^2    - \bar{h}_2\Bb^2  \Ab_1 \Bb^1$
plus their hermitian conjugates, and similarly for the $B$ fields. However, being us interested in the quartic terms of the effective action, we can safely approximate the EOM as in (\ref{EOM1}).}
\beq
\label{EOM1}
D^2 A^i = D^2 B_i = 0 \qquad , \qquad \Db^2 \bar{A}_i = \Db^2 \bar{B}^i = 0
\eeq
from which further useful equations follow
\beq
\label{EOM2}
i \pa^{\a\b} D_\b A^i = i \pa^{\a\b} D_\b B_i  = 0 \qquad , \qquad
i \pa^{\a\b} \Db_\b \bar{A}_i = i \pa^{\a\b} \Db_\b \bar{B}^i  = 0
\eeq
In principle, setting the external superfields on--shell might cause problems when IR divergences appear in loop integrals. We dimensionally regularize these divergences working in $D=3-2\e$ dimensions, while keeping spinors and $\e_{ijk}$ tensors strictly in three dimensions.  We then use the prescription to set the external superfields on--shell at finite $\e$.

To summarize, the general strategy is the following:  For a given process and at a given order in loops we draw all super--Feynman diagrams with four external scalar superfields. The corresponding contribution will be the product of a color/combinatorial factor times a function of the kinematic variables.
We work in the large $M,N$ limit and perturbatively in $\l = M/K_1$ and $\hat{\l} = N/K_2$.
To determine the kinematic function, we perform D--algebra to reduce superdiagrams to a linear combination of ordinary momentum integrals. This is achieved by integrating by parts spinorial derivatives coming from vertices and propagators and using the algebra (\ref{algebra}) up to the stage where only one factor $D^2 \Db^2$ for each loop is left.
This procedure is highly simplified by the on--shell conditions (\ref{EOM1}, \ref{EOM2}) on the external superfields. We then evaluate momentum integrals in dimensional regularization by using standard techniques (Feynman parametrization
and Mellin-Barnes integrals).

In momentum space the external superfields carry outgoing momenta $(p_1, p_2, p_3, p_4)$, with $p_i^2=0$ and
$\sum_i p_i =0$. At the level of the effective action we are allowed to conveniently rename the external momenta, since the $p_i$'s are integrated.
When evaluating the amplitude, the total contribution from every single graph will be given by the sum over all possible permutations of the external legs accounting for the different scattering channels.

Mandelstam variables are defined as $s=(p_1+p_2)^2, t=(p_1+p_4)^2, u=(p_1+p_3)^2$.

\subsection{Scattering at one--loop}

For a generic ${\cal N}=2$ model described by the action (\ref{action}), we first concentrate on the chiral amplitudes $(A^i B_j A^k B_l)$.

\begin{figure}[h!]
    \centering
    \includegraphics[width=0.4\textwidth]{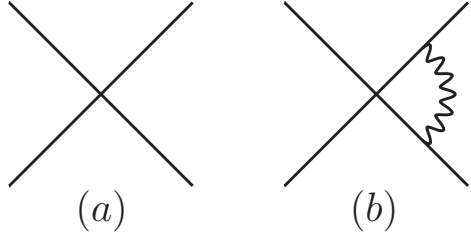}
    \caption{Diagrams contributing to the tree level and one--loop four--point chiral scattering amplitude. }
    \label{0set}
\end{figure}

At tree level and one loop the corresponding contributions are depicted in Fig. \ref{0set} where the four--point interaction comes from the superpotential term in (\ref{action}).

The tree--level amplitudes as coming from Fig. \ref{0set}$(a)$ are simply given by
\bea
&& {\cal A}_4^{tree}(A^1(p_1), B_1(p_2), A^2(p_3), B_2 (p_4)) =  h_1
\non \\
&& {\cal A}_4^{tree}(A^1(p_1), B_2(p_2), A^2(p_3), B_1 (p_4)) =  h_2
\eea
At one loop, we need evaluate diagram \ref{0set}$(b)$.  Performing on--shell D--algebra and going to momentum space, the corresponding term in the effective action turns out to be proportional to
\bea
\label{1loopchiral}
&& \int \,d^4\theta  \; {\Tr (A^i(p_1)  B_j(p_2) D^\a A^k(p_3) D^\b B_l(p_4) ) }\times
\intke{k} \frac{(k + p_4)_{\a\b}}{k^2 (k-p_3)^2 (k+p_4)^2}
\non \\
&&\xrightarrow{\e \rightarrow 0}  \frac18 \, \int \,d^4\theta \; {\Tr (A^i(p_1)  B_j(p_2) D^\a A^k(p_3) D^\b B_l(p_4) ) }\;
\frac{(p_4 - p_3)_{\a\b}}{|p_3 + p_4|^3}
\eea
where in the second line we have used the results (\ref{triangle}, \ref{vector-triangle}) for the scalar and vector--like triangles in dimensional regularization.

Now, using the on--shellness conditions (\ref{EOM2}), which in the case under exam read
$p_3^{\a\b} D_\a A^k(p_3)=0$ and $p_4^{\a\b} D_\b B_l(p_4)=0$, it is easy to see that the final result is zero. Since the same pattern occurs for all the permutations of the external momenta, we conclude that
the chiral four--point amplitude is one--loop vanishing. This occurs not only in the planar limit, but also for any finite value of $M,N$.

Notably, the one--loop vanishing of the effective action, quartic in the chiral superfields, can be proved to be true even off--shell \cite{AndreaMati}.

\vskip 10pt
We now consider non--chiral amplitudes of the type $(A^i \bar{A}_j A^k \bar{A}_l)$. For generic
${\cal N}=2$ superCFT's we do not expect them to be related to the chiral amplitudes. Therefore, a priori there is no reason to expect them to vanish.

The relevant diagrams for these amplitudes are listed in Fig. \ref{fig:twoloops}.

\FIGURE{
  \centering
\includegraphics[width = 0.9 \textwidth]{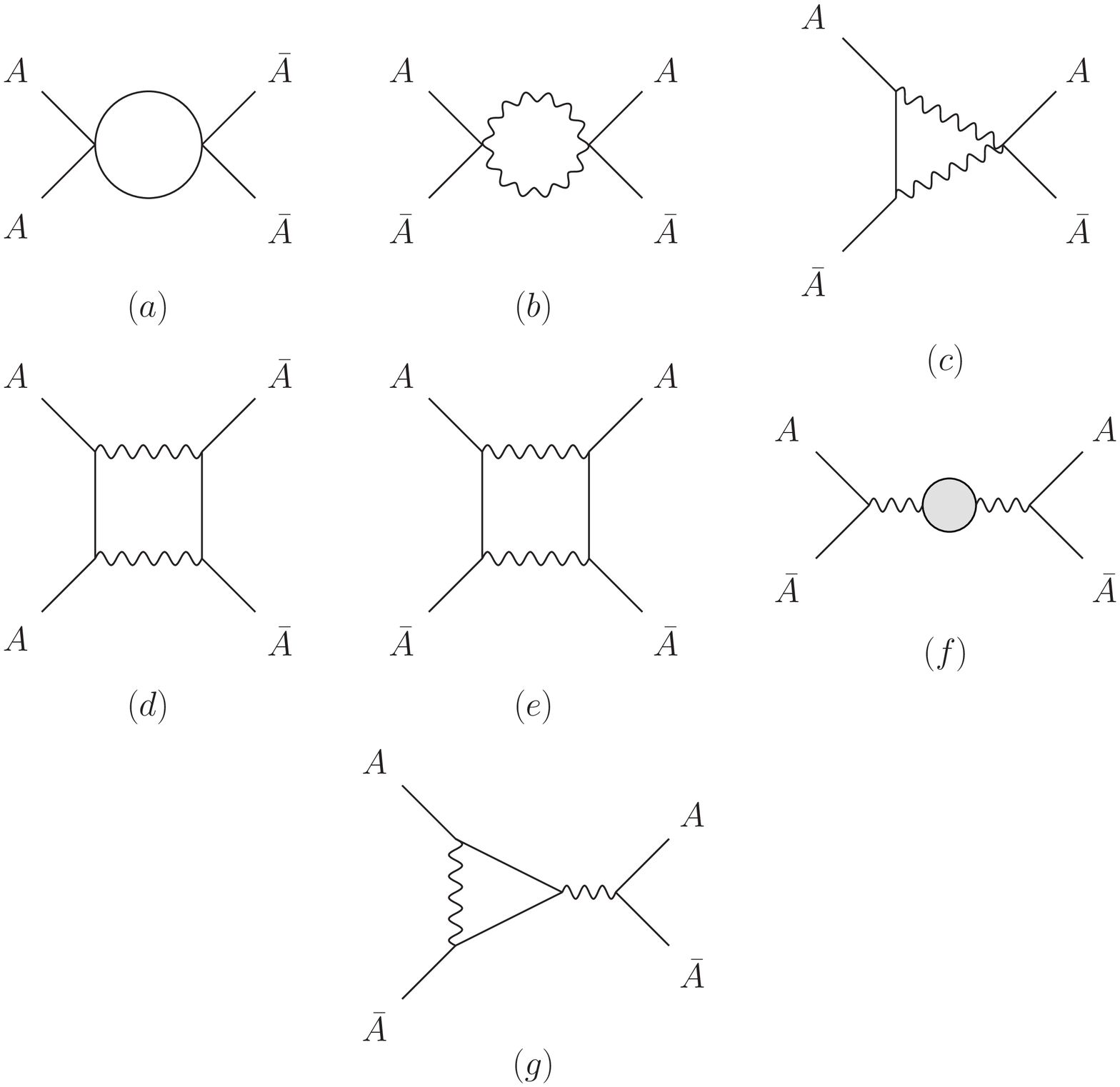}
\caption{One--loop diagrams contributing to non--chiral amplitudes $(A^i \bar{A}_j A^k \bar{A}_l)$.}
\label{fig:twoloops}
}

For each graph we compute the corresponding color/combinatorial factor and perform on--shell D--algebra.  We list the results valid for $M,N$ finite (for the time being, no large $M,N$ limit is taken). Since we work at the level of the effective action, an overall integral over $p_i$ momenta is understood. We also neglect an overall $(4\pi)^2$ coming from the gauge propagators (\ref{gaugeprop}).

\vskip 7pt
\noindent
\underline{Diagram \ref{fig:twoloops}$(a)$ }:
This is the only diagram which involves the chiral interaction vertices proportional to $h_1,h_2$. In this case D--algebra is trivial and the resulting color structure gives only double traces. Exploiting the possibility to relabel the integrated momenta, the result can be written in a quite compact form
\bea
2(a) =
\frac{1}{32\pi^2} \, \int \,d^4\theta \; && \!\!\!\!\Big\{  (|h_1|^2   + |h_2|^2 ) \;
\Tr (A^i(p_1) \Ab_i(p_2))  \, \Tr (A^j (p_4) \Ab_j (p_3) )
 \\&&
+  (h_1 \hb_2 + h_2 \hb_1) \; \Tr (A^i(p_1) \Ab_j(p_3))  \, \Tr (A^j (p_4) \Ab_i (p_2) )
\non\\&&
-  |h_1 + h_2|^2 \, \Tr (A^i(p_1) \Ab_i(p_2))  \, \Tr (A^i(p_4) \Ab_i (p_3) ) \Big\} \; {\cal B}(p_1+p_4)
\non
\eea
where ${\cal B}(p_1+p_4)$ is the bubble integral defined in (\ref{bubble}). Repeated indices are understood to be summed.

\vskip 7pt
\noindent
\underline{Diagram \ref{fig:twoloops}$(b)$ }: In this case the result is a linear combination of single and double traces. Single traces are associated with planar graphs and are leading in the large $M,N$ limit.
D--algebra is easily performed and leads to
\bea
2(b) = \int \,d^4\theta \; \Bigg\{ &-&
\frac{M}{4 K_1^2} \; \Tr (A^i(p_1) \Ab_i(p_2) A^j (p_4) \Ab_j (p_3) )
\\
&-& \frac{N}{4 K_2^2} \; \Tr (A^i(p_1) \Ab_j(p_3) A^j (p_4) \Ab_i (p_2) )
\non\\
&-& \left(  \frac{1}{4 K_1^2} + \frac{1}{4 K_2^2}
\right)  \; \Tr (A^i(p_1) \Ab_i(p_2)) \,  \Tr (A^j (p_4) \Ab_j (p_3) )
\non\\
&-& \frac{1}{ K_1 K_2}  \; \Tr (A^i(p_1) \Ab_j(p_3)) \,  \Tr (A^j (p_4) \Ab_i (p_2) )
\Bigg\}   \; {\cal B} (p_1+p_2)
\non
\eea

\vskip 7pt
\noindent
\underline{Diagram \ref{fig:twoloops}$(c)$ }: With a convenient choice for the internal momentum, this diagram gives rise to
\bea
2(c) = \int \,d^4\theta \; && \!\!\!\!\Bigg\{
 \frac{M}{2 K_1^2}  \; \Tr ( D^{\a} A^i (p_1) \Db^{\b} \Ab_i (p_2) A^j (p_4) \Ab_j (p_3))
\non\\
&&+
 \frac{N}{2 K_2^2} \; \Tr ( D^{\a} A^i (p_1) \Ab_j (p_3) A^j (p_4) \Db^{\b} \Ab_i (p_2))
  \non \\
 &&+
  \left( \frac{1}{2 K_1^2} + \frac{1}{2 K_2^2} \right) \;
\Tr ( D^{\a} A^i (p_1) \Db^{\b} \Ab_i (p_2) ) \,  \Tr( A^j (p_4) \Ab_j (p_3))
\non\\
&&+
\frac{2}{K_1 K_2}  \; \Tr ( D^{\a} A^i (p_1) \Ab_j (p_3) )  \, \Tr( A^j (p_4) \Db^{\b} \Ab_i (p_2))
 \Bigg\}
 \non\\&& \times \intke{k}\frac{k_{\a\b}}{k^2 (k-p_1)^2 (k+p_2)^2}
\eea
where D--algebra requires integrating two spinorial derivatives on the external fields.

Using the result (\ref{vector-triangle}) for the vector--like triangle in dimensional regularization, this contribution vanishes due to the equations of motion (\ref{EOM2}) $p_1^{\a\b} D_\a A(p_1)=0$ and $p_2^{\a\b} \Db_\b \Ab(p_2)=0$.

\vskip 7pt
\noindent
\underline{Diagram \ref{fig:twoloops}$(d)$ }: This is the first case where on--shell D--algebra
and repeated use of the equations of motion  allow for a drastic simplification
of the final result. We give few details of the calculation.

Computing the color factors we obtain only double trace structures. After performing D--algebra
we are led to
\bea
\int \,d^4\theta && \; \left\{  \left(\frac{1}{2K_1^2} +\frac{1}{2K_2^2} \right) \;
\Tr \left(D^\a  A^i(p_1) \Db^\b \Ab_i(p_2)\right) \,  \Tr \left( D^\g A^j (p_4) \Db^\d \Ab_j (p_3) \right)  \right.
\non\\&& \left.
~ - \frac{1}{K_1 K_2}  \; \Tr \left( D^\a A^i(p_1) \Db^\d \Ab_j(p_3)\right) \,  \Tr \left( D^\g A^j (p_4) \Db^\b \Ab_i (p_2) \right)
\right\}
\non\\&&
\times \intke{k} \frac{ k_{\a\g} k_{\b\d} }{k^2 (k+p_1)^2 (k+p_1+p_4)^2 (k-p_2)^2}
\eea
We first integrate by parts the $\Db^\b$ derivative. Using the equations of motions (\ref{EOM1},\ref{EOM2})
we obtain
\bea
\label{partial}
\int \,d^4\theta && \; \left\{  \left(\frac{1}{2K_1^2} +\frac{1}{2K_2^2} \right) \;
\Tr \left(A^i(p_1)  \Ab_i(p_2)\right) \,  \Tr \left( D^\g A^j (p_4) \Db^\d \Ab_j (p_3) \right)  \right.
\non\\&& \left.
~ - \frac{1}{K_1 K_2}  \; \Tr \left(  A^i(p_1) \Db^\d \Ab_j(p_3)\right) \,  \Tr \left( D^\g A^j (p_4)  \Ab_i (p_2) \right)
\right\}
\non\\&&
\times \intke{k} \frac{ p_1^{\a\b} k_{\a\g} k_{\b\d} }{k^2 (k+p_1)^2 (k+p_1+p_4)^2 (k-p_2)^2}
\non \\
\non \\
&& - \left\{  \left(\frac{1}{2K_1^2} +\frac{1}{2K_2^2} \right) \;
\Tr \left(D^\a  A^i(p_1) \Ab_i(p_2)\right) \,  \Tr \left( A^j (p_4) \Db^\d \Ab_j (p_3) \right)  \right.
\non\\&& \left.
~ + \frac{1}{K_1 K_2}  \; \Tr \left( D^\a A^i(p_1) \Db^\d \Ab_j(p_3)\right) \,  \Tr \left( A^j (p_4) \Ab_i (p_2) \right)
\right\}
\non\\&&
\times \intke{k} \frac{  p_4^{\b\g} k_{\a\g} k_{\b\d} }{k^2 (k+p_1)^2 (k+p_1+p_4)^2 (k-p_2)^2}
\eea
We concentrate on the first integral. The numerator can be rewritten as
\beq
p_1^{\a\b} k_{\a\g} k_{\b\d} = p_1^{\a\b} \left[ k_{\d\g} k_{\b\a} - k^2  C_{\d\a} C_{\b\g}  \right] =
(k+p_1)^2  k_{\g\d} - k^2 (k+p_1)_{\g\d}
\eeq
Now, simplifying the squares at numerator against the ones at denominators we are left with a linear combination of scalar and vector--like  triangle integrals. Exploiting the fact that in dimensional regularization the scalar triangle is zero, while the vector--like one is proportional to a bubble integral (see Appendix B), the first term in (\ref{partial})  reduces to
\bea
\int \,d^4\theta &&  \left\{  \left(\frac{1}{2K_1^2} +\frac{1}{2K_2^2} \right) \;
\Tr \left(A^i(p_1)  \Ab_i(p_2)\right) \,  \Tr \left( D^\g A^j (p_4) \Db^\d \Ab_j (p_3) \right)  \right.
\\&& \left.
 - \frac{1}{K_1 K_2}  \; \Tr \left(  A^i(p_1) \Db^\d \Ab_j(p_3)\right) \,  \Tr \left( D^\g A^j (p_4)  \Ab_i (p_2) \right) \right\}
\times \frac{(p_2)_{\g\d}}{(p_1+p_4)^2} \, {\cal B}(p_1 + p_4)
\non
\eea
where equations of motion and momentum conservation have been used. Now, integrating by parts the $\Db^\d$ derivative and using on--shell conditions, it can be further simplified to
\bea
\int \,d^4\theta && \; \left\{  \left(\frac{1}{2K_1^2} +\frac{1}{2K_2^2} \right) \;
\Tr \left(A^i(p_1)  \Ab_i(p_2)\right) \,  \Tr \left(  A^j (p_4) \Ab_j (p_3) \right)  \right.
\\&& \left.
~ + \frac{1}{K_1 K_2}  \; \Tr \left(  A^i(p_1)   \Ab_j(p_3)\right) \,  \Tr \left(   A^j (p_4)  \Ab_i (p_2) \right) \right\}
\times \frac{(p_2+p_4)^2}{(p_1+p_4)^2} \, {\cal B}(p_1 + p_4)
\non
\eea
We can apply the same tricks to the second integral in eq. (\ref{partial}). After a bit of algebra, we obtain a similar expression which, summed to the rest, leads to the final expression for the box diagram $2(d)$ in terms of a linear combination of bubbles
\bea
2(d) = &&\int \,d^4\theta \;  \left\{  \left(\frac{1}{2K_1^2} +\frac{1}{2K_2^2} \right) \;
\Tr \left( A^i(p_1) \Ab_i(p_2)\right) \,  \Tr \left( A^j (p_4) \Ab_j (p_3) \right)  \right.
\\&& \left.
~+ \frac{1}{K_1 K_2}  \; \Tr \left( A^i(p_1) \Ab_j(p_3)\right) \,  \Tr \left( A^j (p_4) \Ab_i (p_2) \right)
\right\}
 \times
\left[ {\cal B}(p_1+p_2) - {\cal B}(p_1+p_4) \right]
\non
\eea

\vskip 7pt
\noindent
\underline{Diagram \ref{fig:twoloops}$(e)$ }: The result for this diagram reads
\bea
2(e) = -  \int \,d^4\theta \; && \!\!\!\!\Bigg\{
 \frac{M}{2K_1^2} \; \Tr \left(A^i(p_1) \Ab_i(p_2)  D^{\a} A^j (p_4)  \Db^{\b} \Ab_j (p_3) \right)
 \\&&
- \frac{N}{2K_2^2} \; \Tr \left(A^i(p_1)  \Db^{\b} \Ab_j(p_3)  D^{\a} A^j (p_4) \Ab_i (p_2) \right)
 \non\\&&
- \frac{1}{K_1 K_2} \; \Tr \left( A^i(p_1)  \Db^{\b} \Ab_j(p_3)\right) \,  \Tr \left(\Ab_i (p_2)  D^{\a} A^j (p_4) \right)
\Bigg\}
\non\\&&
\times \, (p_2)_{\a}^{\,\,\g} \,  \intke{k} \frac{ k_{\g}^{\,\,\d} \, (k+p_1+p_3)_{\d\b} }{k^2 (k+p_1)^2 (k+p_1+p_3)^2 (k-p_2)^2}
\non
\eea
Elaborating its numerator, the integral can be rewritten as
\beq
\intke{k} \frac{ k^2 C_{\b\g}  +  k_{\g}^{\,\,\d} (p_1+p_3)_{\d\b} }{k^2 (k+p_1)^2 (k+p_1+p_3)^2 (k-p_2)^2}
= C_{\b\g} \, {\cal T}(p_3,p_4) + (p_1+p_3)_{\d\b}  \,  {{\cal Q}_{\cal V}}_{\g}^{\, \, \d}
\eeq
As proved in Appendix B, the triangle and vector--like box integrals are ${\cal O}(\e)$ in dimensional regularization (see eqs. (\ref{triangle}, \ref{vector-box})). Therefore, this diagram can be discarded when $\e \to 0$.

\vskip 7pt
\noindent
\underline{Diagram \ref{fig:twoloops}$(f)$ }: We now consider 1P--reducible diagrams.  For graph
\ref{fig:twoloops}$(f)$ using the one--loop correction to the gauge propagator given in eq. (\ref{1loopprop}) we obtain
\bea
\label{partialf}
\int d^4\theta && \Bigg\{ \frac{1}{K_1^2} \left( N - \frac{M}{4} \right)
\Tr \left( D^\a A^i(p_1) \Ab_i(p_2) A^j (p_4) \Db^\b \Ab_j (p_3) \right)
\\&&
~+ \frac{1}{K_2^2} \left( M - \frac{N}{4} \right) \; \Tr \left(\Ab_i(p_2) D^\a A^i(p_1) \Db^\b \Ab_j (p_3) A^j (p_4)\right)
\non\\&&
~+  \left( \frac{1}{4\,K_1^2}  + \frac{1}{4\,K_2^2} + \frac{2}{K_1 K_2} \right) \; \Tr \left( D^\a A^i(p_1) \Ab_i(p_2)\right) \Tr\left(A^j (p_4) \Db^\b \Ab_j (p_3) \right)  \Bigg\}
\non \\
&&  \qquad\times \frac{(p_4)_{\a\b}}{(p_1+p_2)^2}
{\cal B}(p_1+p_2)
\non
\eea
We can integrate by parts the $D^\a$ derivative. Exploiting the equations of motion, the only non--vanishing term is the one where the derivative hits $\Db^\b \Ab_j (p_3)$, giving a factor $- p_3^{\a\b} (p_4)_{\a\b} =
- (p_3+p_4)^2$. By momentum conservation, this cancels against the denominator in (\ref{partialf})
and we finally obtain
\bea
2(f) &=& - \int  d^4\theta \Bigg\{ \frac{1}{K_1^2} \left( N - \frac{M}{4} \right)
\Tr \left( A^i(p_1) \Ab_i(p_2) A^j (p_4) \Ab_j (p_3) \right)
\\&&
\!\! + \frac{1}{K_2^2} \left( M - \frac{N}{4} \right) \; \Tr \left(\Ab_i(p_2) A^i(p_1) \Ab_j (p_3) A^j (p_4)\right)
\non\\&&
\!\! +  \left( \frac{1}{4\,K_1^2}  + \frac{1}{4\,K_2^2} + \frac{2}{K_1 K_2} \right) \! \Tr \left( A^i(p_1) \Ab_i(p_2)\right) \Tr\left( A^j (p_4) \Ab_j (p_3) \right)  \Bigg\}   \times   {\cal B}(p_1+p_2)
\non
\eea

\vskip 7pt
\noindent
\underline{Diagram \ref{fig:twoloops}$(g)$ }: Finally, we consider the reducible triangle diagram. Performing on--shell D--algebra we produce terms with four spinorial derivatives acting on the external fields. After integrating by parts one of these derivatives, using on--shell conditions and momentum conservation and relabeling internal and external momenta,  we can write the result as
\bea
\label{partialg}
- \int \,d^4\theta\!\!&& \Bigg\{
\frac{N}{K_1 K_2}  \; \Tr \left(A^i(p_1) \Db^\b \Ab_i(p_2) D^\a A^j (p_4) \Ab_j (p_3) \right)
\\&&
-  \frac{M}{K_1 K_2}  \; \Tr \left(A^i(p_1) \Ab_j(p_3) D^\a A^j (p_4) \Db^\b \Ab_i (p_2) \right)
\non\\&&
+  \left(\frac{1}{K_1^2} + \frac{1}{k_2^2}\right) \; \Tr \left( A^i(p_1) \Db^\b \Ab_i(p_2)\right) \,
\Tr \left( D^\a A^j (p_4) \Ab_j (p_3) \right) \Bigg\}
\non\\&&
\times \, \frac{(p_1+p_2)_{\g\a}}{(p_1+p_2)^2} \, \intke{k} \frac{ (k+p_1)^{\g\d} \, k_{\d\b}}{ k^2 (k+p_1)^2 (k-p_2)^2 }
\non
\eea
We can elaborate the numerator of the integrand to obtain
\bea
 \intke{k} \frac{ k^2 \d^\g_{\, \b}  +p_1^{\g\d} \, k_{\d\b}}{ k^2 (k+p_1)^2 (k-p_2)^2 }  &=& \d^\g_{\, \b}  \, {\cal B}(p_1+p_2) + p_1^{\g\d} \, {{\cal T}_{\cal V}}_{\d\b}
\non \\
&=&    {\cal B}(p_1+p_2) \left[ \d^\g_{\, \b} -  \frac{p_1^{\g\d} (p_2)_{\d\b}}{(p_1+p_2)^2} \right]
\eea
where eq. (\ref{vector-triangle}) has been used together with on--shell conditions. Inserting back into
eq. (\ref{partialg}), observing that on--shell $(p_1+p_2)_{\g\a} p_1^{\g\d} (p_2)_{\d\b} = (p_1+p_2)^2
(p_2)_{\a\b}$ where $(p_2)_{\a\b}$ vanishes when  contracted with $\Db^\b \Ab_i (p_2)$, we obtain
\bea
- \int \,d^4\theta\!\!&& \Bigg\{
\frac{N}{K_1 K_2}  \; \Tr \left(A^i(p_1) \Db^\b \Ab_i(p_2) D^\a A^j (p_4) \Ab_j (p_3) \right)
\non \\
&&
-  \frac{M}{K_1 K_2}  \; \Tr \left(A^i(p_1) \Ab_j(p_3) D^\a A^j (p_4) \Db^\b \Ab_i (p_2) \right)
\non\\&&
+  \left(\frac{1}{K_1^2} + \frac{1}{K_2^2}\right) \; \Tr \left( A^i(p_1) \Db^\b \Ab_i(p_2)\right) \,
\Tr \left( D^\a A^j (p_4) \Ab_j (p_3) \right) \Bigg\}
\non\\&&
\times \, \frac{(p_1)_{\a\b}}{(p_1+p_2)^2}  \, {\cal B}(p_1+p_2)
\eea
On--shell conditions are once again helpful for reducing the structure of spinorial derivatives acting on the external fields. In fact, integrating by parts the $D^\a$ derivative we produce a term $p_2^{\a\b}$ that, contracted with $(p_1)_{\a\b}$, cancels $(p_1 +p_2)^2$ at the denominator. We finally obtain
\bea
2(g) = - \int \,d^4\theta\!\!&& \Bigg\{
\frac{N}{K_1 K_2}  \; \Tr \left(A^i(p_1) \Ab_i(p_2) A^j (p_4) \Ab_j (p_3) \right)
\\&&
+  \frac{M}{K_1 K_2}  \; \Tr \left(A^i(p_1) \Ab_j(p_3) A^j (p_4) \Ab_i (p_2) \right)
\non\\&&
+  \left(\frac{1}{K_1^2} + \frac{1}{K_2^2}\right) \; \Tr \left( A^i(p_1) \Ab_i(p_2)\right) \,
\Tr \left( A^j (p_4) \Ab_j (p_3) \right) \Bigg\}  \times  {\cal B}(p_1+p_2)
\non
\eea

We are now ready to sum all the results and obtain the one--loop effective action needed for the evaluation of $(A^i \bar{A}_j A^k \bar{A}_l)$ amplitudes.

Having reduced all the expressions to strings of external superfields with no derivatives acting on them, multiplied by bubble integrals, we can group them according to their trace structure. We have
single trace contributions from diagrams $2(b), (f), (g)$ and double trace contributions from  $2(a), (b), (d), (f), (g)$.  Collecting them all, we obtain
\bea
\label{1loopfinal}
\int \,d^4\theta\!\!\!\!&& \Bigg\{  - {\Tr (A^i(p_1) \Ab_i(p_2) A^j (p_4) \Ab_j (p_3) )}  \;
\hat{\l}  \left( \frac{1}{K_1} + \frac{1}{K_2} \right) \times {\cal B}(p_1+p_2)
\non \\
&~&~~
- {\Tr (A^i(p_1) \Ab_j (p_3) A^j (p_4) \Ab_i(p_2) )}  \;  \l \,  \left( \frac{1}{K_1} + \frac{1}{K_2} \right)
 \times {\cal B}(p_1+p_2)
\non \\
\non \\
 \\
&&
~~ + {\Tr (A^i(p_1) \Ab_i(p_2)) \,  \Tr (A^j (p_4) \Ab_j (p_3) )} \times
\non \\
&~& \quad \Bigg[ \frac{1}{2}
\left(  \frac{|h_1|^2 + |h_2|^2}{16\pi^2}   - \frac{1}{K_1^2} - \frac{1}{K_2^2}  \right)  \times {\cal B}(p_1+p_4)
-\left( \frac{1}{K_1} + \frac{1}{K_2} \right)^2  \times {\cal B}(p_1+p_2) \Bigg]
\non\\
&&~~ + {\Tr (A^i(p_1) \Ab_j(p_3)) \,  \Tr (A^j (p_4) \Ab_i (p_2) )} \left( \frac{h_1\hb_2 + h_2\hb_1}{32\pi^2}
-\frac{1}{K_1 K_2}  \right) \times {\cal B}(p_1+p_4)
\non \\
 &&~~ -   \frac{1}{32\pi^2}    \; {\Tr (A^i(p_1) \Ab_i(p_2)) \,  \Tr (A^i (p_4) \Ab_i (p_3) )} \; \left| h_1 + h_2 \right|^2
 \times {\cal B}(p_1+p_4)  \Bigg\}
 \non
\eea
First of all, we observe that for $M,N$ finite the quartic effective action, and consequently the amplitude, vanishes when $K_2=-K_1$ and $h_2 = - h_1 = 4\pi/K$. This is exactly the ${\cal N}=6$ superconformal fixed point corresponding to the ABJ theory. This result was expected and provides a non--trivial check of our calculation. In fact, in the ABJ model the non--chiral amplitude is related to the chiral one by $SU(4)$ symmetry and we have already checked that the chiral amplitude is one--loop vanishing.

Taking $M,N$ large and assuming the $h_i$ couplings  of order of $1/K_i$, only single trace contributions survive in (\ref{1loopfinal}). In this case, the amplitude will vanish for $K_2 = -K_1$, independently of the values of the chiral couplings. In particular, we have a vanishing non--chiral amplitude for the whole set of ${\cal N}=2$ superCFT's given by the condition (see eq. (\ref{U(1)}))
\beq
|h_1|^2 + |h_2|^2 = \frac{32\pi^2}{K^2}
\eeq
We observe that the amplitude never vanishes for theories with $K_2 \neq-K_1$, in particular for superCFT's which correspond to turning on a Romans mass in the dual supergravity background.

The same pattern occurs for the $(B_i \bar{B}^j B_k \bar{B}^l)$ amplitudes. In fact, repeating the previous calculation we obtain exactly the same expression (\ref{1loopfinal}) as a consequence of the $Z_2$ symmetry of the action under exchange $V \leftrightarrow \hat{V}$, $A \leftrightarrow B$, $M \leftrightarrow
N$, $K_1 \leftrightarrow K_2$ and $h_1 \leftrightarrow h_2$.

Finally, we need consider mixed amplitudes of the type $(A^i \bar{A}_j \bar{B}^k B_l)$. The contributing diagrams are still the ones drawn in Fig. (\ref{fig:twoloops}) with obvious substitution of one $(A, \bar{A})$ couple with a $(B, \bar{B})$ couple. Applying the same procedure as before, we obtain
\bea
\int \,d^4\theta\!\!&& \Bigg\{ 2 \, {\Tr (A^i(p_1) \Ab_i(p_2) \Bb^j (p_3) B_j (p_4) )} \; \hat{\l} \,  \left(
\frac{1}{K_1} + \frac{1}{K_2} \right) \times {\cal B}(p_1+p_2)
\\
&& ~+ 2 \, {\Tr (\Ab_i(p_2) A^i(p_1) B_j (p_4) \Bb^j (p_3) ) }\, \l \,  \left( \frac{1}{K_1} + \frac{1}{K_2} \right) \times {\cal B}(p_1+p_2)
\non \\
&& ~+ \, {\Tr (A^i(p_1) \Ab_i(p_2) \Bb^j (p_3) B_j (p_4) ) }_{i \neq j} \; M \left( \frac{|h_1|^2}{16\pi^2}
- \frac{1}{K_1^2} \right) \times {\cal B}(p_1+p_4)
\non \\
&& ~+ \, {\Tr (A^i(p_1) \Ab_i(p_2) \Bb^i(p_3) B_i (p_4) )}  \; M \left(
 \frac{|h_2|^2}{16\pi^2}  - \frac{1}{K_1^2} \right) \times {\cal B}(p_1+p_4)
\non \\
&& ~+ \, {\Tr (\Ab_i(p_2) A^i(p_1) B_j (p_4) \Bb^j (p_3) )}_{i \neq j} \; N \left(
\frac{|h_2|^2}{16\pi^2}  -\frac{1}{K_2^2}  \right) \times {\cal B}(p_1+p_4)
\non \\
&& ~+ \, {\Tr (\Ab_i(p_2) A^i(p_1) B_i (p_4) \Bb^i (p_3) )}   \; N \left(
\frac{|h_1|^2}{16\pi^2}  -\frac{1}{K_2^2}  \right) \times {\cal B}(p_1+p_4)
\non \\
\non \\
&&
+ \, 2 \, {\Tr (A^i(p_1) \Ab_i(p_2)) \,  \Tr (B_j (p_4) \Bb^j (p_3) )} \, \left( \frac{1}{K_1} + \frac{1}{K_2} \right)^2 \times {\cal B}(p_1+p_2)
\non \\
&& + \, 2 \, {\Tr (A^i(p_1) B_j(p_4)) \,  \Tr (\Ab_i (p_2) \Bb^j (p_3) )} \left( \frac{ h_1\hb_2 + h_2\hb_1}{32\pi^2}  -\frac{1}{K_1 K_2} \right) \times {\cal B}(p_1+p_4)  \Bigg\}
\non
\eea
For $M,N$ finite, these amplitudes vanish only at the ABJ(M) fixed point (\ref{ABJ}), as expected. However, in contrast with the previous case, in the large $M,N$ limit a non--trivial dependence on the chiral couplings survives, which restricts the set of superCFT's with vanishing one--loop amplitudes to be only the ABJ(M) models.

In the ABJM case, this result is consistent with what has been found in Ref. \cite{ABM} in components and massive regularization, and in \cite{CH} by means of the generalized unitarity cuts method.

\subsection{Scattering at two loops}

We restrict to the ABJ model (\ref{ABJ}) for which we have found that all the four--point amplitudes vanish at one loop. This result is consistent with the one--loop vanishing of the Wilson loop and leads to conjecture the existence of a WL/amplitudes duality for this theory. To give evidence to this conjecture, we evaluate
four--point scattering amplitudes at two loops.

We study amplitudes of the type $(A^i B_j A^k B_l)$, where the external $A,B$ particles carry outgoing momenta $p_1,\dots,p_4$ ($p_i^2=0$).

At two loops, in the planar sector, the amplitude can be read from the single trace part of the two--loop effective superpotential
\begin{eqnarray}
\label {effepotential}
&&\Gamma^{(2)}[A,B]=\int d^2\theta d^3p_1\dots d^3p_4\,(2\pi)^3\,\delta^{(3)}({\sum}_i p_i)\times\non\\
&&\frac{2\pi }{K}\epsilon_{ik}\epsilon^{jl}\,\mathrm{tr}\left(A^i(p_1) B_j(p_2)A^k(p_3)B_l(p_4)\right)\sum\limits_{X=a}^{g}\mathcal{M}^{(X)}(p_1,\dots,p_4)
\end{eqnarray}
where the sum runs over the six 1PI diagrams in Fig. \ref{1set}, plus the contribution from the 1P--reducible (1PR) graph in Fig. \ref{1set}(g) where the bubble indicates the two--loop correction to the chiral propagator.

In (\ref{effepotential}) we have factorized the tree level expression, so that
$ \mathcal{M}^{(X)}(p_1,\dots,p_4)$ are contributions to ${\cal A}_4/{\cal A}_4^{tree}$.

In order to evaluate the diagrams we fix the convention for the upper--left leg
to carry outgoing momentum $p_1$ and name the other legs counterclockwise.
The momentum--dependent contributions in (\ref{effepotential}) are the product of a combinatorial factor times a sum of ordinary Feynman momentum integrals arising after performing D--algebra on each supergraph (more details can be found in \cite{AndreaMati}). There are a total of four diagrams of the classes $(b)$, $(c)$, $(f)$ and $(g)$, eight diagrams of the classes $(d)$ and $(e)$ and two diagrams of the class $(a)$. The color/flavor factors $\mathcal{C}_i$ for them are given by
\begin{align}
&\mathcal{C}_a=(4\pi)^2\,\tfrac{\lambda^2+\hat\lambda^2}{2}\quad\
\mathcal{C}_b=(4\pi)^2\,\tfrac{\lambda^2+\hat\lambda^2}{8}\quad\
\mathcal{C}_c=(4\pi)^2\,\tfrac{8\lambda\hat\lambda-\lambda^2-\hat\lambda^2}{2}\nonumber\\
&\mathcal{C}_d=(4\pi)^2\,\tfrac{\lambda^2+\hat\lambda^2}{4}\quad\
\mathcal{C}_e=(4\pi)^2\,\lambda\hat\lambda\quad\
\mathcal{C}_f=-(4\pi)^2\,\lambda\hat\lambda
\end{align}
while diagram $(g)$ contains subdiagrams with different color/flavor factors and these cannot be factorized.

\FIGURE{
    \centering
    \includegraphics[width=0.7\textwidth]{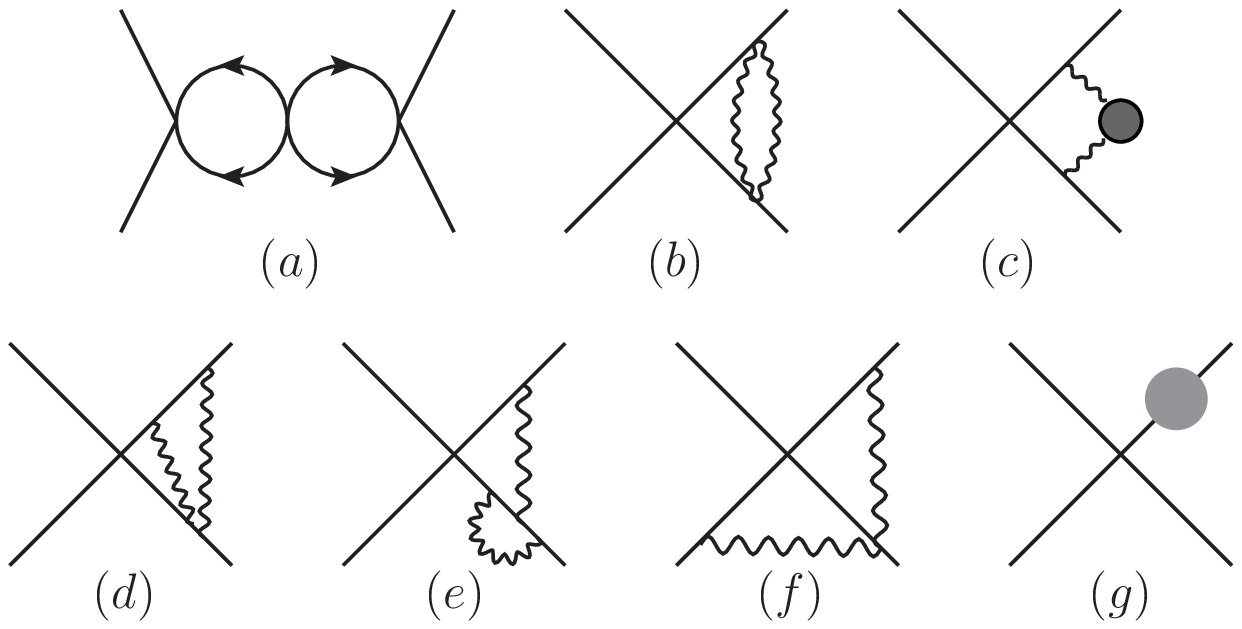}
    \caption{Diagrams contributing to the two--loop four--point scattering amplitude. The dark--gray blob represents one--loop corrections and the light--gray blob two--loop ones.}
    \label{1set}
}

\vskip 7pt
\noindent
\underline{Diagram \ref{1set}$(a)$ }: We begin with the simplest graph which,
after D--algebra, reduces to the following factorized Feynman integral
\begin{equation}\label{DalgebraA}
\mathcal{D}^s_a=\mu^{4\epsilon}\int\frac{d^d k}{(2\pi)^d}\frac{d^d l}{(2\pi)^d}\frac{-(p_1+p_2)^2}{k^2\,(k+p_1+p_2)^2\, l^2\,(l-p_3-p_4)^2}
\end{equation}
where $\mu$ is the mass scale of dimensional regularization.

The $k$ and the $l$ bubble integrals can be separately evaluated using the result (\ref{bubble}), so obtaining
\begin{equation}
\mathcal{D}^s_a =
-G[1,1]^2\left(\frac{\mu^2}{s}\right)^{2\epsilon}
\end{equation}
In order to obtain the corresponding contribution to the amplitude, we need sum over all the independent configurations of the external momenta. Inserting the corresponding color/flavor factors we obtain
\begin{equation}
\label{a}
\mathcal{M}^{(a)}=-8\pi^2(\lambda^2+\hat{\lambda}^2) G[1,1]^2\left(\left(\frac{\mu^2}{s}\right)^{2\epsilon}+\left(\frac{\mu^2}{t}\right)^{2\epsilon}\right)=
-\frac{3}{2}\zeta_2(\lambda^2+\hat{\lambda}^2)+\mathcal{O}(\epsilon)
\end{equation}

\vskip 7pt
\noindent
\underline{Diagram \ref{1set}$(b)$ }: After D--algebra,  it gives
\begin{equation}\label{DalgebraB}
\mathcal{D}^{s_1}_b=\mu^{4\epsilon}\int\frac{d^d k}{(2\pi)^d}\frac{d^d l}{(2\pi)^d}\frac{2(p_3+p_4)^2}{l^2\,(l+k)^2\,(k-p_4)^2\,(k+p_3)^2}
\end{equation}
Performing the $l$ integral with the help of Eq. (\ref{bubble}) , we obtain a triangle integral with a modified exponent in one of its propagators
\begin{equation}
\mathcal{D}^{s_1}_b=\mu^{4\epsilon}\,G[1,1]\,\int\frac{d^d k}{(2\pi)^d}\frac{2(p_3+p_4)^2}{(k^2)^{1/2+\epsilon}\,(k-p_4)^2\,(k+p_3)^2}
\end{equation}
We Feynman--parametrize the denominator and integrate over the momentum $k$. Taking into account that we are working on--shell ($p_i^2=0$) we finally obtain
\bea
\label{bpartial}
\mathcal{D}^{s_1}_b &=& \frac{\mu^{4\epsilon}\,2s\,G[1,1]\,\Gamma(1+2\epsilon)}{(4\pi)^{d/2}\Gamma(1/2+\epsilon)}
\int\limits_0^1 \frac{d\beta_1 d\beta_2 d\beta_3\,\delta(\sum_i \beta_i-1)\beta_1^{-1/2+\epsilon}}{( \beta_1\beta_2 p_4^2+\beta_1\beta_3 p_3^2+\beta_2\beta_3 s)^{1+2\epsilon}}
\non \\
&\xrightarrow{p_{3,4}^2 \rightarrow 0} &
\frac{2 G[1,1] \Gamma(1+2\epsilon)\Gamma^2(-2\epsilon)}{(4\pi)^{d/2}\Gamma(1/2-3\epsilon)}
\left(\frac{\mu^2}{s}\right)^{2\epsilon}
\eea
where $\Gamma^2(-2\epsilon)$ signals the presence of an on--shell IR divergence.

Summing over the four inequivalent configurations of the external legs multiplied by the correct vertex factors, the contribution to the amplitude reads
\begin{align}
\label{b}
& \mathcal{M}^{(b)}= \, 8\pi^2 (\lambda^2+\hat{\lambda}^2) \frac{ G[1,1] \Gamma(1+2\epsilon)\Gamma^2(-2\epsilon)}{(4\pi)^{d/2}\Gamma(1/2-3\epsilon)}
\left(\left(\frac{\mu^2}{s}\right)^{2\epsilon}+\left(\frac{\mu^2}{t}\right)^{2\epsilon}\right)\nonumber\\
= & \,\frac{\lambda^2+\hat\lambda^2}{8}\,\left[
\frac{1}{(2\epsilon)^2}\left(\frac{s}{\pi e^{-\gamma_E}\mu^2}\right)^{-2\epsilon}+
\frac{1}{(2\epsilon)^2}\left(\frac{t}{\pi e^{-\gamma_E}\mu^2}\right)^{-2\epsilon}\,-\,\frac{5}{2}\,\zeta_2\,+
\,\mathcal{O}(\epsilon)\right]
\end{align}

\vskip 7pt
\noindent
\underline{Diagram \ref{1set}$(c)$ }: This diagram may result problematic, being infrared divergent even when evaluated off--shell.  In fact, after D--algebra, the particular diagram drawn in Fig. \ref{1set}$(c)$ gives rise to the following integral
\begin{equation}\label{DalgebraC}
\mathcal{D}^{s_1}_c=\frac{\mu^{4\epsilon}}{2}\int\frac{d^d k}{(2\pi)^d}\frac{d^d l}{(2\pi)^d}
\frac{Tr(\gamma_{\mu}\gamma_{\nu}\gamma_{\rho}\gamma_{\sigma})\,
p_3^{\mu}\,(k-p_3)^{\nu}\,(k+p_4)^{\rho}\,p_4^{\sigma}}
{l^2\,(l+k)^2\,k^2\,(k-p_4)^2\,(k+p_3)^2}
\end{equation}
which, performing the bubble $l$ integral and using the identity
\begin{equation}
Tr(\gamma_{\mu}\gamma_{\nu}\gamma_{\rho}\gamma_{\sigma})\,
p_3^{\mu}\,(k-p_3)^{\nu}\,(k+p_4)^{\rho}\,p_4^{\sigma}=(p_3+p_4)^2 k^2-p_3^2(k+p_4)^2-p_4^2(k-p_3)^2
\end{equation}
can be separated into three pieces
\begin{equation}\label{DalgebraC2}
\mathcal{D}^{s_1}_c=\frac{1}{4}\,\mathcal{D}^{s_1}_b-\frac{1}{2}G[1,1]G[1,3/2+\epsilon]\,(p_3^2)^{-2\epsilon}
-\frac{1}{2}G[1,1]G[1,3/2+\epsilon]\,(p_4^2)^{-2\epsilon}
\end{equation}
While the first term is the off--shell well--behaving Feynman integral in Eq. (\ref{bpartial}) that in the on--shell limit produces $1/\e$ poles, the second and third terms are badly divergent even off--shell. However, we can show that these unphysical divergences  are cured when we add the 1PR diagrams corresponding to two--loop self--energy corrections to the superpotential, as depicted in Fig. \ref{1set}$(g)$.

In fact, for example the contribution from the diagram with the two--loop correction on the fourth leg as drawn in the picture, yields
\begin{equation}\label{gamba}
\mathcal{D}^{4}_g=-8\pi^2(8\lambda\hat{\lambda}-\lambda^2-\hat{\lambda}^2)
\,G[1,1]\,G[1,3/2+\epsilon]\,(p_4^2)^{-2\epsilon}+32\pi^2\lambda\hat{\lambda}\; p_4^2 \, {\cal B}(p_4)^2
\end{equation}
where color factors have been included.

It is easy to realize that the first term of this expression is off--shell infrared divergent, but precisely  cancels the third term in (\ref{DalgebraC2}) when all the vertex factors of diagram \ref{1set}$(c)$ are taken into account.  On the other hand, the second term in (\ref{gamba}) comes from a double factorized bubble which vanishes on--shell.

Since in a similar way the second term in (\ref{DalgebraC2}) gets canceled by a diagram with a
two--loop correction on the third leg, summing diagrams \ref{1set}$(c)$,$(g)$ and their permutations
we are finally led to the following interesting identity
\begin{equation}
\label{c}
\mathcal{M}^{(c)}+\mathcal{M}^{(g)}=\frac{\lambda^2+\hat{\lambda}^2-8\lambda\hat{\lambda}}{\lambda^2+\hat{\lambda}^2}
\,\mathcal{M}^{(b)}
\end{equation}

\vskip 7pt
\noindent
\underline{Diagram \ref{1set}$(d)$ }: Diagrams of type $(d)$ may be calculated using Mellin-Barnes techniques. Specifically, after D--algebra the  diagram in figure  gives rise to the integral
\begin{equation}\label{DalgebraD}
\mathcal{D}^{s1}_d =\mu^{4\epsilon}\int \frac{d^d k}{(2\pi)^d}\frac{d^d l}{(2\pi)^d}
\frac{Tr(\gamma_{\mu}\gamma_{\nu}\gamma_{\rho}\gamma_{\sigma})\,
p_4^{\mu}\,(p_3+p_4)^{\nu}\,(k+p_4)^{\rho}\,(l-p_4)^{\sigma}}{(k+p_4)^2\,(k-p_3)^2\,(k+l)^2\,(l-p_4)^2\,l^2}
\end{equation}
Using the identity (\ref{MBvectortriangle}) and the on--shell conditions,  it can be rewritten as
\begin{align}
\mathcal{D}_d^{s1} = &\frac{-s\,\Gamma(1/2-\epsilon)}{(4\pi)^{d/2}\Gamma(1-2\epsilon)}
\int\limits_{-i\infty}^{i\infty}\frac{d{\bf z}}{2\pi i}\Gamma(-{\bf z}|1+{\bf z}|3/2+\epsilon+{\bf z}|-1/2-\epsilon-
{\bf z})
\non  \\
& \times\;  \mu^{4\e} \, \int\frac{d^d k}{(2\pi)^d}\frac{1}{(k^2)^{3/2+\epsilon+{\bf z}}\,[(k+p_4)^2]^{-{\bf z}}\,(k-p_3)^2}
\end{align}
The integral over $k$ can be easily performed by Feynman parametrization, leading to
\bea
\mathcal{D}_d^{s1} &=&-\mu^{4\e} \, \frac{\Gamma(1/2-\!\epsilon|1+\!2\epsilon|-\!2\epsilon)}
{(4\pi)^d\Gamma(1-\!2\epsilon|1/2-\!3\epsilon)}
\non \\
&~& \times \,
\int\limits_{-i\infty}^{i\infty}\frac{d{\bf z}}{2\pi i}\,\Gamma(1+{\bf z}|3/2+\epsilon+{\bf z}|-1/2+\epsilon-{\bf z}|-1-2\epsilon-{\bf z})
\non \\
&=&-\frac{\Gamma^3(1/2-\epsilon)\Gamma(1+2\epsilon)\Gamma^2(-2\epsilon)}
{(4\pi)^{d}\Gamma^2(1-2\epsilon)\Gamma(1/2-3\epsilon)\,(s/\mu^2)^{2\epsilon}}
\eea
where the remaining integral over the complex variable ${\bf z}$ has been performed by using the Barnes first Lemma (\ref{lemma}).

Taking into account the eight permutations with corresponding flavor/color factors we obtain
\begin{align}
\label{d}
&\mathcal{M}^{(d)} =\,
-16\pi^2(\lambda^2+\hat{\lambda}^2)\frac{\Gamma^3(1/2-\epsilon)\Gamma(1+2\epsilon)\Gamma^2(-2\epsilon)}
{(4\pi)^{d}\Gamma^2(1-2\epsilon)\Gamma(1/2-3\epsilon)}\left(\left(\frac{\mu^2}{s}\right)^{2\epsilon}+ \left(\frac{\mu^2}{t}\right)^{2\epsilon}\right)\nonumber\\
= & \,\frac{\lambda^2+\hat\lambda^2}{4}\,\left[
-\frac{1}{(2\epsilon)^2}\left(\frac{s}{4\pi e^{-\gamma_E}\mu^2}\right)^{-2\epsilon}-
\frac{1}{(2\epsilon)^2}\left(\frac{t}{4\pi e^{-\gamma_E}\mu^2}\right)^{-2\epsilon}\,+\,\frac{7}{2}\,\zeta_2\,+
\,\mathcal{O}(\epsilon)\right]
\end{align}

\vskip 7pt
\noindent
\underline{Diagram \ref{1set}$(e)$ }: Using the identities derived in \cite{AndreaMati} it is possible to write this diagram as a combination of diagrams $(b)$ and $(d)$. It holds that
\begin{equation}
(1+\mathcal{S}_{34})\mathcal{D}_e^{s1}=(1+\mathcal{S}_{34})\mathcal{D}_d^{s1}+\mathcal{D}_b^{s1}-p_3^2\mathcal{B}(p_3)^2
-p_4^2\mathcal{B}(p_4)^2
\end{equation}
where $\mathcal{D}_e^{s1}$ is the particular diagram of type $(e)$ drawn in the figure and the operator $(1+\mathcal{S}_{34})$ symmetrizes the diagram with respect to the third and fourth leg. Notice the presence of a double factorized bubble which can be dropped on--shell. Accounting for all permutations and flavor/color factors we thus find that
\begin{equation}
\label{e}
\mathcal{M}^{(e)}= 4\frac{\lambda\hat{\lambda}}{\lambda^2+\hat{\lambda}^2}
\left(\mathcal{M}^{(d)}+2\,\mathcal{M}^{(b)}\right)
\end{equation}

\vskip 7pt
\noindent
\underline{Diagram \ref{1set}$(f)$ }: The most complicated contribution comes from this diagram, as it involves a non--trivial function of the $s/t$ ratio. Surprisingly, after some cancelations it turns out to be finite.

The D--algebra for the specific choice of external momenta as in figure results in the Feynman integral
\begin{equation}\label{DalgebraE}
\mathcal{D}^{234}_{f}=\mu^{4\epsilon}\int \frac{d^d k}{(2\pi)^d}\frac{d^d l}{(2\pi)^d}
\frac{-Tr(\gamma_{\mu}\gamma_{\nu}\gamma_{\rho}\gamma_{\sigma})\,
p_4^{\mu}\,p_2^{\nu}\,k^{\rho}\,l^{\sigma}}{k^2\,(k-p_2)^2\,(k+l+p_3)^2\,(l-p_4)^2\,l^2}
\end{equation}
Again, using Eq. (\ref{MBvectortriangle}) for the $k$ integral and working on-shell, we obtain
\begin{align}\label{DalgebraE2}
\mathcal{D}^{234}_{f} &=&
\frac{\Gamma(1/2-\epsilon)}{(4\pi)^d\Gamma(1-2\epsilon)}\int\limits_{-i\infty}^{i\infty}\frac{d{\bf z}}{2\pi i}
\Gamma(-{\bf z})\Gamma(1+{\bf z})\Gamma(3/2+\epsilon+{\bf z})\Gamma(-1/2-\epsilon-{\bf z})
\nonumber\\
&~& \times \; \mu^{4\e} \int\frac{d^d l}{(2\pi)^d}\,
\frac{Tr(\gamma_{\mu}\gamma_{\nu}\gamma_{\rho}\gamma_{\sigma})\,p_4^{\mu}\,p_2^{\nu}\,(l+p_3)^{\rho}\,l^{\sigma}}
{l^2\,(l-p_4)^2\,[(l+p_3)^2]^{-{\bf z}}\,[(l+p_2+p_3)^2]^{3/2+\epsilon+{\bf z}}}
\end{align}
It is convenient to separate the $l$ integral in two pieces by using the on--shell identity
\begin{equation}\label{DalgebraE3}
Tr(\gamma_{\mu}\gamma_{\nu}\gamma_{\rho}\gamma_{\sigma})\,p_4^{\mu}\,p_2^{\nu}\,(l+p_3)^{\rho}\,l^{\sigma}
\Big{|}_{\mbox{\tiny{on--shell}}}=
-(s+t)\,l^2+
Tr(\gamma_{\mu}\gamma_{\nu}\gamma_{\rho}\gamma_{\sigma})\,p_4^{\mu}\,p_2^{\nu}\,p_3^{\rho}\,l^{\sigma}
\end{equation}
The first piece in (\ref{DalgebraE3}) contains an $l^2$ factor which cancels the $l^2$ propagator in (\ref{DalgebraE2}) leading to a simple triangle which is straightforwardly evaluated as we did for the previous diagram.

The second piece is a vector--box integral which after Feynman parametrization can be written in terms of a second 1--fold Mellin-Barnes integral. Interchanging the order of the two Mellin-Barnes integrals and solving for the original one in (\ref{DalgebraE2}) with the first Barnes lemma (\ref{lemma}), the total result is
\begin{align}\label{DalgebraE4}
&\mathcal{D}^{234}_{f} =
\frac{(s+t)\Gamma^3(1/2-\epsilon)\mu^{4\epsilon}}
{(4\pi)^d\Gamma(1/2-3\epsilon)\Gamma^2(1-2\epsilon)}
\left[ -\phantom{\int\limits_{-i\infty}^{i\infty}}\!\!\!\!\!\!
\frac{\Gamma(1+2\epsilon)\Gamma^2(-2\epsilon)}{s^{1+2\epsilon}}\,\,+\right.\nonumber\\
&+\left.\frac{1}
{t^{1+2\epsilon}}
\int\limits_{-i\infty}^{i\infty}\!\frac{d{\bf v}}{2\pi i}
\Gamma(-{\bf v})\Gamma(-2\epsilon\!-\!{\bf v})\Gamma(-1\!-2\epsilon\!-\!{\bf v})
\Gamma^2(1\!+\!{\bf v})\Gamma(2\!+\!2\epsilon+\!{\bf v})\left(\frac{s}{t}\right)^{{\bf v}}\right]
\end{align}
The contour of the Mellin-Barnes integral in the second term of the last expression is not well--defined in the limit $\epsilon\to 0$, reflecting the presence of poles in $\epsilon$. The reason is that in this limit the first pole of $\Gamma(-1-2\epsilon-\!{\bf v})$ collapses with the first pole of $\Gamma^2(1+\!{\bf v})$.
In order to have a well defined contour in the $\e \to 0 $ limit, we can deform the contour so that it passes on the right of the point ${\bf v}=-1-2\epsilon$ and include the residue of the integrand in this point. Surprisingly, it turns out that this residue exactly cancels the first term in (\ref{DalgebraE4}) so that we obtain a simple 1--fold Mellin-Barnes integral which is finite in the limit $\epsilon\to 0$
\begin{align}
&\mathcal{D}^{234}_{f} =\frac{(1+s/t)\Gamma^3(1/2-\epsilon)}{(4\pi)^d\Gamma^2(1-2\epsilon)\Gamma(1/2-3\epsilon)(t/\mu^2)^{2\epsilon}} \\
\times\int\limits^{+i\infty}_{-i\infty} \frac{d{\bf v}}{2\pi i}\Gamma(-{\bf v}) & \Gamma(-2\epsilon -{\bf v})
\Gamma^{*}(-1-2\epsilon-{\bf v})\Gamma^2(1+{\bf v})\Gamma(2+2\epsilon+{\bf v})\left(\frac{s}{t}\right)^{{\bf v}}
\end{align}
This integral can be calculated in the $\epsilon\to 0$ limit by closing the contour and performing the infinite sum of all the residues inside it. Taking into account all four permutations of the diagram and flavor/color factors we finally obtain
\begin{equation}
\label{f}
\mathcal{M}^{(f)}=\lambda\hat{\lambda}\left(\tfrac{1}{2}\ln^2(s/t)+3\zeta_2\right)+\mathcal{O}(\epsilon)
\end{equation}

\vskip 10pt
We are now ready to collect the partial results (\ref{a}, \ref{b}, \ref{c}, \ref{d}, \ref{e}, \ref{f}) and find the four--point chiral superamplitude at two loops. After some algebra, and redefining the mass scale as
\beq
\mu'^2=  2^{\sigma^2/2} \, (8  \pi  e^{-\gamma_E}\,\mu^2 ) 
\eeq
the result can be cast into the following compact form
\begin{equation}
\label{result}
\framebox[14 cm][c]{$
 \displaystyle \mathcal{M}^{(2)} \equiv \frac{\mathcal{A}_4^{(2 \, loops)}}{\mathcal{A}^{tree}_4} = \bar\lambda^2\, \left[-\frac{( s/\mu'^2)^{-2\epsilon}}{(2\, \epsilon)^2}-\frac{(t/\mu'^2)^{-2\epsilon}}{(2\, \epsilon)^2}+\frac12\,\ln^2 \left(\frac{s}{t}\right)+ C_{\cal A} +
\mathcal{O}(\epsilon)\right]
 $}
\end{equation}
where $\bar{\l} = \sqrt{\l \hat{\l}} = \sqrt{MN}/k$ and $C_{\cal A}$  is a constant given by
\begin{equation}
C_{\cal A} =\left(4-\tfrac{5}{4}\sigma^2\right)\zeta_2+
\left(1+\tfrac{1}{2}\sigma^2\right)\left(3+\tfrac{1}{2}\sigma^2\right)\ln^2 2
\end{equation}
We note that the mass scale and the constant depend non--trivially on the parity--violating parameter $\s$ defined in eq. (\ref{sigma}). Since it only appears as a square, parity is not violated at this stage.
As a check, we observe that for $\sigma= 0$ the result reduces to the ABJM amplitudes computed in
\cite{BLMPS1}.

\section{Discussion}

We now discuss the main properties of our result (\ref{result}) for the four--point amplitude at two loops.

First of all, in the ABJM case ($\sigma= 0$) the result coincides with the one in \cite{CH} obtained by applying generalized unitarity methods. In particular, the effective mass scale is the same and the analytical expression for the constant $C_{\cal A} |_{\sigma\to 0}=4\zeta_2+3\ln^2 2$ matches the numerical result of \cite{CH}.

In Ref. \cite{CH} the result has been found by assuming {\em a priori} that in the planar limit dual conformal invariance should work also at quantum level. In fact, an ansatz has been made on the general structure of the amplitude which turns out to be a linear combination of integrals that, if extended off--shell, are well defined in three dimensions and exhibit conformal invariance in the dual $x$--space ($p_i = x_i - x_{i+1}$).

On the other hand, our calculation relies on a standard Feynman diagram approach which does not make use of any assumption. The identification of the two results is then a remarkable proof of the validity of on--shell dual conformal invariance for this kind of theories.

\subsection{Amplitudes/WL duality}

In the general ABJ case, if we write the Mandelstam variables in terms of the dual ones, $s = x_{13}^2$ and $t=x_{24}^2$, up to a (scheme--dependent) constant our result matches those in Eqs.
(\ref{resultWL1}, \ref{WL2result}) for the two--loop expansion of a light--like Wilson loop, once we formally identify the IR and UV rescaled regulators of the scattering amplitude and the Wilson loops  as $\mu'^2= 1/\mu_{WL}'^2$ or $ \mu'^2= 1/\mu_{WL}''^{\, 2}$.

Since  the Wilson loop is conformally invariant in the ordinary configuration space, the identification of
the two--loop amplitude with the corresponding term in the WL expansion is a further proof of dual conformal invariance in the on--shell sector of the theory.

We remind that the two results (\ref{resultWL1}, \ref{WL2result}) in Section 3 correspond to two
possible definitions of Wilson loop in ABJ models.
At this stage, the result (\ref{result}) seems to match both. However, if we compare the rescaled mass regulators, we see that apart from the sign of the Euler constant, in the result for the amplitude $\mu'$ looks like $\mu'_{WL}$ in eq. (\ref{rescaling1}), whereas it is quite different from $\mu''_{WL}$  in Eq. (\ref{rescaling2}). Although there is no particular reason for the two mass scales to match exactly, this might be a first indication that the definition (\ref{WL1}) for the light--like Wilson loop dual to scattering amplitudes is preferable.

\subsection{Dual conformal invariance}

As for the ${\cal N}=4$ SYM case, in the ABJ models  the two--loop on--shell amplitude divided by its tree--level contribution, when written in terms of dual variables has the same functional structure as the second order expansion of a light--like Wilson loop.
Wilson loops are invariant under the transformations of the standard conformal group of the ABJ theory, even though UV divergences break this symmetry anomalously. Hence, the on--shell amplitude should inherit this symmetry, possibly anomalously broken by IR divergences.

In fact, in the ${\cal N}=4$ SYM case where the amplitudes/WL duality also works, the perturbative results for planar MHV scattering amplitudes divided by their tree--level contribution can be expressed as linear combinations of scalar integrals that are off--shell finite and dual conformally invariant \cite{Drummond:2006rz, Drummond:2007aua} in four dimensions.  Precisely, once written in terms of dual variables, the integrands times the measure are invariant under translations, rotations, dilatations and special conformal transformations.
Dual conformal invariance is broken on--shell by IR divergences that require introducing a mass regulator. Therefore, conformal Ward identities acquire an anomalous contribution \cite{Drummond:2007au}.

A natural consequence of our findings is that the two--loop result for three dimensional ABJ(M) models should also exhibit dual conformal invariance, and then it should be possible to rewrite the final expression (\ref{result}) for the on--shell amplitude as a linear combination of scalar integrals which are off--shell finite in three dimensions and manifestly dual conformally invariant at the level of the integrands. Indeed, for the ABJM case this has been proved in Ref. \cite{CH} where the amplitude has been obtained by unitarity cuts method, based on the ansatz for the amplitude to be dual conformal invariant.

Following \cite{CH}, we introduce a set of independent scalar integrals $I_{1s}, I_{2s}, I_{3s}, I_{5s} \equiv I_{1s}-I_{4s}$ which correspond to the following off--shell, three dimensional, dual conformally invariant expressions
\bea
\label{dual1}
&& I_{1s} = \int \frac{d^3 x_5 d^3 x_6}{(2\pi)^6} \; \frac{x_{13}^4}{x_{15}^2 x_{35}^2 x_{56}^2 x_{16}^2 x_{36}^2}
\\ \label{dual2}
&& I_{2s} = \int \frac{d^3 x_5 d^3 x_6}{(2\pi)^6} \; \frac{x_{13}^2 x_{24}^2}{x_{15}^2 x_{35}^2 x_{45}^2 x_{16}^2 x_{26}^2 x_{36}^2}
\\ \label{dual3}
&& I_{3s} = \int \frac{d^3 x_5 d^3 x_6}{(2\pi)^6} \; \frac{x_{13}^2 x_{24}^2}{x_{35}^2 x_{45}^2 x_{56}^2  x_{26}^2 x_{16}^2}
\\ \label{dual4}
&& I_{4s} = \int \frac{d^3 x_5 d^3 x_6}{(2\pi)^6} \; \frac{x_{13}^2 x_{25}^2 x_{46}^2}{x_{15}^2 x_{35}^2 x_{45}^2 x_{56}^2 x_{16}^2 x_{26}^2 x_{36}^2}
\eea
plus their $t$--counterparts obtained by cyclic permutation of the $(1,2,3,4)$ indices.
Their graphical representation is given in Fig. \ref{dualinvariant}.

\FIGURE{
    \centering
    \includegraphics[width=0.7\textwidth]{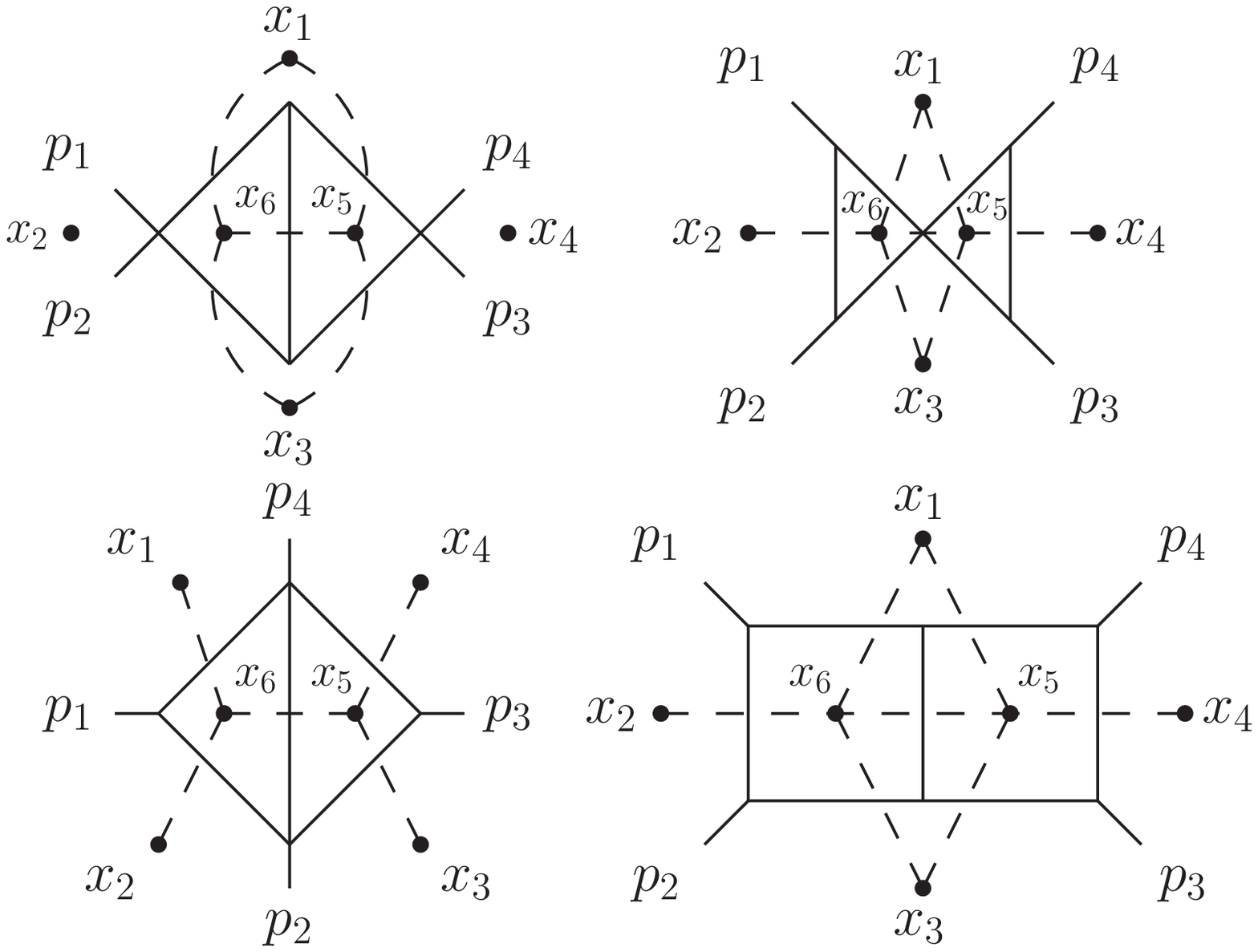}
    \caption{Graphical representation of dual conformally invariant integrals.}
    \label{dualinvariant}
}

The appearance of the particular combination $I_{1s} - I_{4s}$ is not an accident. In fact,
due to the presence of internal cubic vertices, the integrals $I_{1s}$, $I_{4s}$ are IR divergent also
off--shell and then ill--defined in three dimensions. Dual conformal invariance would require to discharge these integrals. However, as we show in Appendix C, taking the linear combination $I_{1s}-I_{4s}$
the off--shell divergences  cancel and $I_{5s}$ is well--defined in three dimensions.
 
The on--shell evaluation of these integrals in $D=3-2\e$ dimensions reveals that 
\beq
I_{2s} \sim O(\e^2) \qquad \qquad I_{3s} + I_{3t} = - I_{1s} - I_{1t}
\eeq
Thus, the actual basis for the four--point scattering amplitude reduces to $I_{1s}, I_{5s}, I_{1t}, I_{5t}$.  Evaluating them and defining $\bar{\mu}^2 = 8\pi e^{-\g_E} \mu^2$, one finds \cite{CH} \footnote{In \cite{CH}, the constant part of $(I_{5s} + I_{5t})$ has been evaluated numerically. However, {\em a posteriori} one can check that the numerical factor is well reproduced by the analytical expression in (\ref{I5}). }  
\bea
\label{I1}
&& I_{1s} + I_{1t} =  - \frac{1}{16\pi^2} \left[ \frac{\left( s/\bar{\mu}^2 \right)^{-2\e}}{2\e}   +  
\frac{\left( t/\bar{\mu}^2 \right)^{-2\e}}{2\e}    + 2 - 2\ln{2}   \right] + O(\e) 
\\  
&& I_{5s} + I_{5t} = - \frac{1}{8\pi^2} \left[ \frac{\left( s/\bar{\mu}^2 \right)^{-2\e}}{(2\e)^2}   +  
\frac{\left( t/\bar{\mu}^2 \right)^{-2\e}}{(2\e)^2}  - \frac{\left( s/\bar{\mu}^2 \right)^{-2\e}}{2\e}   - 
\frac{\left( t/\bar{\mu}^2 \right)^{-2\e}}{2\e} \right.
\non \\ \label{I5}
&~& \qquad \qquad \qquad \quad \quad \left.  - \frac12 \ln^2{ \left( \frac{s}{t} \right)} + 2\ln{2} -2 - 
3 \ln^2{2} - 4 \z_2 \right] + O(\e) 
\eea
Using these results, it is easy to see that for the ABJM theory, the two--loop amplitude (\ref{result}) for $\s = 0$ can be written as
\beq
\label{result2}
\mathcal{M}^{(2)}_4 \Big|_{\rm ABJM}= (4\pi \lambda)^2 \left[ \frac12 I_{5s} +  I_{1s} + ( s \leftrightarrow t) \right]
\eeq
Remarkably, this linear combination not only reproduces correctly the non--trivial part of the amplitude, but also fits the numerical constant $C_{\cal A}$.  In particular, it is such that the non--maximal transcendentality terms in (\ref{I1}), (\ref{I5}) cancel.

We can now generalize this analysis to the ABJ models where the amplitude has the same functional structure of the ABJM one, except for a non--trivial dependence on the $\s$ parameter in the mass scale and in the $C_{\cal A}$ constant. Because of the appearance of $\s$, we find that in terms of the integrals (\ref{I1}, \ref{I5}) given as functions of the $\bar\mu^2$ scale, the 
$\mathcal{M}^{(2)}_4 $ ratio can be written as
\beq
\label{result3}
\mathcal{M}^{(2)}_4  \Big|_{\rm ABJ} = (4\pi \bar\lambda)^2 \left[ \frac12 I_{5s}(\bar\mu^2) + \left(1 + \frac{\s^2}{2} \ln{2} \right) I_{1s}(\bar\mu^2) + ( s \leftrightarrow t) \right]  + C_{\rm res}   
\eeq
where $C_{\rm res}$ is a residual constant given by
\beq
C_{\rm res} =  \bar\lambda^2 \, \s^2 \left( \ln^2{2} - \frac{5}{4} \z_2 + \ln{2} \right)
\eeq
The non--trivial appearance of $\s^2$ in the coefficients might reflect the fact that in the ABJ case one cannot factorize completely the color dependence outside the combination of integrals. This would suggest that a generalization of the unitarity cuts method should be employed where the trace structures are not stripped out. Nevertheless, it is not difficult to see that the application of such a method would never reproduce the $\ln{2}$ coefficient in front of $I_{1s}, I_{1t}$. 

One would be tempted to conclude that ABJ amplitudes cannot be computed by unitarity cuts methods. However, a way out
is to start from a linear combination of (\ref{I1}, \ref{I5}) integrals where the mass parameter has been rescaled as $\bar\mu^2 \to 
\mu'^2 = \bar\mu^2 2^{\s^2/2}$. Doing that, we find that the two loop ratio can now be written as
\beq
\label{result4}
\mathcal{M}^{(2)}_4 \Big|_{\rm ABJ}= (4\pi \bar\lambda)^2 \left[ \frac12 I_{5s}(\mu'^2) +  I_{1s}(\mu'^2) + ( s \leftrightarrow t) 
  \right] + C'_{\rm res} 
\eeq
where 
\beq
C'_{\rm res} = \bar{\l}^2 \s^2 \left[ \left( 2 + \frac{\s^2}{4} \right) \ln^2{2} - \frac54 \z_2 \right] 
\eeq
The situation has drastically improved, since rational coefficients in front of the integrals indicate that the same result could be obtained by unitarity cuts method.  However, in that approach the question of why and how fixing {\em a priori} a non--standard mass scale in the dual invariant integrals remains an open problem.

Except for the particular case $M=N$, in general the basis of scalar integrals selected by dual conformal symmetry reproduces the four--point amplitude only up to a constant. 
This is a quite different result compared to what happens in the ABJM and ${\cal N}=4$ SYM cases where dual conformal integrals reproduce exactly the four--point amplitude. However, at the order we are working, 
the difference is only by a constant and dual conformal invariance is safe, as well as the anomalous Ward identities which follow.

At higher loops, we expect the non--trivial dependence on $\sigma$ to affect also the terms depending on the kinematic variables. It would be very interesting to check whether this phenomenon may spoil dual conformal invariance or higher order amplitudes could still be expressed as ($\sigma$--dependent) combinations of dual conformal invariant integrals. For this reason, it would extremely important to evaluate the amplitude at four loops.

\vskip 10pt
 
In the ABJM theories, comparing the result for the amplitude obtained by ordinary perturbative methods with the one obtained by unitarity cuts method, we can write
\beq
\label{equality}
\mathcal{M}^{(2)}_4 = \sum_i c_i J_i \Big|_{\text{\scriptsize $\begin{array}{c} D=3-2\epsilon \\ on-shell
\end{array}$}}
= \sum_{n} \a_n I_n \Big|_{\text{\scriptsize $\begin{array}{c} D=3-2\epsilon \\ on-shell
\end{array}$}}
\eeq
where $J_i$ are momentum integrals associated to the Feynman diagrams in Fig. \ref{1set}, whereas $I_n$ are the scalar integrals (\ref{dual1}-\ref{dual4}).

Since the linear combination on the right hand side, when written strictly in $D=3$ with $x_{i,i+1}^2 \neq 0$ is dual conformally invariant, the natural question which arises is whether also the left hand side shares the same property.

In order to answer this question, we investigate the behavior of the Feynman integrals $J_i$ under dual conformal transformations, off--shell and in three dimensions.
After rewriting them in terms of dual space variables, we implement the inversion  $x_{ij}^2 \to \frac{x_{ij}^2}{x_i^2 x_j^2}$ and $d^dx_i \to \frac{d^dx_i}{(x_i^2)^d}$, which is the only non--trivial conformal transformation to be checked.

As in four dimensions the invariance under inversion rules out triangle and bubble--like diagrams, similarly in three dimensions it forbids the appearance of bubbles.
Therefore, just looking at the integrands, we see that the integrals associated to diagrams \ref{1set}$(a)$-\ref{1set}$(b)$ cannot be separately dual conformal invariant. Moreover, despite the fact that diagrams  \ref{1set}$(d)$-\ref{1set}$(f)$  consist of triangles only, non--trivial numerators spoil invariance under inversion, as well.

Nevertheless, these considerations on the integrands may fail in very special cases.
As an example, we consider the double bubble diagram \ref{1set}$(a)$. In dual coordinates, the corresponding integral reads
\beq
{\cal B} = x_{13}^2 \, \int \frac{d^3x_5}{x_{15}^2 x_{35}^2 } \, \int  \frac{d^3x_6}{x_{16}^2 x_{36}^2 }
\eeq
Performing inversion, this integral gets mapped into a double triangle integral
\beq
{\cal T} = x_{13}^2 x_1^2 x_3^2 \,  \int \frac{d^3x_5}{x_5^2 x_{15}^2 x_{35}^2 } \, \int  \frac{d^3x_6}{x_6^2x_{16}^2 x_{36}^2 }
\eeq
If we evaluate them off--shell, we obtain ${\cal B} = {\cal T} = 1/64$ (see eqs. (\ref{bubble}, \ref{trioffshell})) . 
Therefore,  the double bubble diagram is invariant under inversion at the level of the integral, even if it is not invariant at the level of the integrand.

Motivated by this example we may wonder whether dual conformal invariance on the left hand side of eq. (\ref{equality}) could be restored at the level of the integrals.
By numerical evaluating the integrals associated to the independent topologies \ref{1set}$(b)$, \ref{1set}$(d)$ and \ref{1set}$(f)$
and to their duals, obtained by acting with conformal inversion, we find that every single integral is not by itself dual conformally invariant.

However, one may still doubt that the situation could improve when summing over all scattering channels, or summing all the contributions to get the total off--shell amplitude.

We find that, even if for every single diagram the sum over permutations of external legs definitively improves the result, as for
a large sample of momentum configurations the integral and its dual look very close to each other, they never appear to be exactly invariant, neither does the total sum.

We thus conclude that the off--shell amplitude computed by Feynman diagrams is not dual conformally invariant.
In other words, the identity (\ref{equality}) is not an algebraic relation between different basis of integrals, as if it were the case it should be valid for any value of the kinematic variables. Instead, it holds only when the integrals are evaluated on the mass--shell and in dimensional regularization. This is not puzzling if we take into account that in three dimensions and in dimensional regularization the on--shell limit is not a smooth limit for the integrals. It would be interesting to investigate what happens when using a different regularization, for example the one suggested in \cite{AHPS, HMN}.

In any case, our result reinforces the statement that dual conformal invariance is a (anomalous) symmetry {\em only} of the on--shell sector of the theory.

\subsection{BDS--like ansatz}

The striking correspondence between the four--point amplitudes of ABJM theory at two loops and the one of 4d $\mathcal{N}=4$ SYM at one loop led us to conjecture \cite{BLMPS1} (see also \cite{CH}) that a BDS--like ansatz \cite{BDS} may be formulated also for the three--dimensional case
\begin{equation}
\label{3dBDS}
\mathcal{M}_4   = e^{Div + \frac{f_{CS}(\l)}{8}\left(\ln^2\left(\frac{s}{t}\right)+ 8\zeta_2 + 6\,\ln^2 2\right)  + C(\lambda) }
\end{equation}
where $f_{CS}(\lambda)$ is the scaling function of ABJM. The analogy between the two theories is due to the fact that they share similar integrable structures, with asymptotic Bethe equations related by an unknown function $h(\lambda)$ \cite{GV}-\cite{Grignani}. This leads to a connection between anomalous dimensions of composite operators and, in particular, to the following relation between the scaling functions \cite{GV}
\begin{equation}\label{3d4d}
f_{CS}(\lambda)= \left. \frac{1}{2}f_{\mathcal{N}=4} (\lambda)\right|_{\frac{\sqrt{\lambda}}{4 \pi}\rightarrow h(\lambda)}
\end{equation}
in terms of the interpolating function $h(\lambda)$ that needs to be determined in perturbation theory. Since the scaling function governs the coefficients of the kinematic part of the four--point amplitude in $\mathcal{N}=4$ SYM by means of the BDS exponentiation, one is tempted to conjecture that an analogue resummation may also hold in the ABJM case, giving rise to equation (\ref{3dBDS}). This formula is confirmed at two loops by the results of \cite{BLMPS1,CH}. 
Since at weak coupling  $h(\lambda)$ is known up to the forth order \cite{MOS1,MOS2,LMMSSST}, we easily find 
\beq
f_{CS}(\lambda)= 4 \lambda^2 - 24\, \zeta_2\, \lambda^4 + {\cal O}(\lambda^6)
\eeq
and the ansatz (\ref{3dBDS}) provides a  prediction for the four--loop expression of the finite remainder $F_4^{(4)}$ (in the notation of  \cite{BDS}) for the ABJM four--point scattering amplitude \cite{BLMPS1} 
\beq
\label{conjecture}
F_4^{(4)} = \frac{\l^4}{8} \ln^4
\left(\frac{s}{t}\right) + \l^4 \left(\frac{3}{2}\ln^2 2- \zeta_2\right) \ln^2\left(\frac{s}{t}\right) + {\rm Consts}
\eeq

Now we discuss how this scenario might be affected by the generalization to the ABJ case, where integrability is not expected to be trivially preserved. We first note that the final expression (\ref{result}) is the result of summing many contributions which in general are proportional to the homogeneous couplings $\l^2$, $\hat{\l}^2$ and to the mixed $\l \hat{\l}$ one. It is interesting to observe that at this order it is possible to redefine $\mu^2$ in such a way that in all the terms depending on the kinematic variables, the contributions proportional to the homogeneous couplings cancel, leading to an expression which is basically identical to the one for ABJM, except for the substitution $N^2 \to MN$. This also happens for the WL computed in Section 2.

This phenomenon has been also observed in the evaluation of the spin--chain Hamiltonian associated to the two--loop anomalous dimension matrix  for single--trace operators \cite{Bak:2008vd} and in the two--loop contribution to the dispersion relation of magnons \cite{MZ}.  This special dependence on the coupling constants is a signal that parity symmetry along with integrability are preserved at least at two--loop order even if ABJ theory is manifestly parity breaking. At four loops only the dispersion relation, i.e. the eigenvalue of the Hamiltonian of spin chains with a single excitation, is known to date. It has been confirmed by explicit computations and the following expansion for the interpolating function has been found \cite{MOS1,MOS2,LMMSSST}
\beq
\label{habjm}
h^2(\bar{\lambda},\sigma)= \bar{\lambda}^2 - \bar{\lambda}^4 \left[4 \zeta_2 + \zeta_2 \, \sigma^2 \right]
\eeq
From the last term of this expression it is clear that, even if the departure from the ABJM case becomes non--trivial, still the function turns out to depend quadratically on $\sigma$ and thus parity breaking is not visible. This might indicate that integrability is not broken also at the four-loop level.

Therefore it seems plausible that, at least up to four loops, the planar limit of ABJ scattering amplitude could behave in the same way as in the ABJM case, being governed by the scaling function $f(\bar{\lambda},\sigma)$ obtained through (\ref{3d4d}) where we 
replace $h(\l)$ with $h(\bar{\lambda},\sigma)$. 

Since  at weak coupling $h(\bar{\lambda},\sigma)$ is known up to four loops \cite{MOS1,MOS2,LMMSSST}, we find 
\beq
f_{CS}(\bar\lambda,\sigma)= 4 \bar\lambda^2 - 4\, (6+\sigma^2)\, \zeta_2\, \bar\lambda^4 + {\cal O}(\bar\lambda^6)
\eeq
It is then easy to see that the four--point amplitude at order $\bar{\l}^2$ can be identified with the first order expansion of 
an exponential of the type in eq. (\ref{3dBDS}) with  $f_{CS}(\bar\lambda,\sigma)= 4 \bar\lambda^2$.

Moreover, this would lead to a prediction for the four-loop expression for the finite remainder of the ABJ four--point amplitude to be given by
\beq
F_4^{(4)} =  \frac{\bar{\lambda}^4}{8} \ln^4
\left(\frac{s}{t}\right) + \bar{\lambda}^4 \left[ \frac{1}{2} \left(1+\tfrac{1}{2}\sigma^2\right)\left(3+\tfrac{1}{2}\sigma^2\right)\ln^2 2- \left(1+ \frac98 \sigma^2\right) \zeta_2  \right] \ln^2\left(\frac{s}{t}\right) + {\rm Consts}
\eeq
At four loops the parity--violating parameter is expected to play an active role and the theory could present a very different behaviour with respect to the ABJM case.
It would be interesting to check this expression by a direct computation.

\subsection{The amplitude at strong coupling}

In ${\cal N}=4$ SYM a recipe for computing scattering amplitudes at strong coupling has been proposed by Alday and Maldacena \cite{AM} within the context of the AdS/CFT correspondence. According to their prescription, the amplitude for $n$ gluons is obtained by computing the minimal area of a surface in the ${\rm AdS}_5$ dual background, ending on a light--like $n$--polygon, whose edges are determined by the gluon momenta
\bea\label{stramp}
{\cal M} = e^{-\frac{R_{AdS_5}^2}{2\p}\, A} + {\cal O}\left( \frac{1}{\sqrt{\l}} \right)
\eea
Here $R_{AdS_5}$ stems for the AdS radius and determines the dependence on the 't Hooft coupling ($R_{AdS_5}^2 =
\sqrt{g^2 N} \equiv \sqrt{\l_{SYM}}$).

After a suitable regularization of this area, the four--gluon amplitude ${\cal M}_4$, to first order in $\sqrt{\l_{SYM}}$, matches exactly the BDS ansatz, where the strong coupling scaling function is plugged in. This provides a remarkable check on the BDS ansatz as well as a hint towards the WL/scattering amplitude duality, since the strong coupling computation of the amplitude strikingly parallels that of a light--like WL.

Motivated by the analogy with the four dimensional case and by evidence in favor of WL/amplitude duality and BDS exponentiation, we investigate the ABJM four--point amplitude at strong coupling, by following the same steps as in \cite{AM}.

At strong coupling where the 't Hooft parameter $\l =N/K$ is large, and in the intermediate regime $K \ll N \ll K^5$ the AdS/CFT correspondence provides a dual description of the ABJM in terms of type IIA supergravity on $AdS_4 \times {\mathbb CP}^3$.

The dual background in string frame is
\bea\label{metricABJM}
ds^2 = \frac{R^3}{K} \left( \frac{1}{4} ds^2_{AdS_4} + ds^2_{\mathbb{CP}^3 } \right)
\eea
where in $l_s$ units the $AdS_4$ radius is given by $R^2_{AdS_4} = \frac{R^3}{4\,K} = \frac{\sqrt{2^5\, \p^2\, K\, N}}{4\,K} = \sqrt{2}\, \p\, \sqrt{\l}$.

A first indication that the general prescription for computing scattering amplitudes at strong coupling could still be (\ref{stramp}) with the parameters conveniently adapted to the three dimensional model, comes from observing that the ratio of the two AdS radii coincides with the ratio of the scaling functions at leading order in the couplings. In fact, taking into account that for ${\cal N}=4$ SYM at strong coupling
\beq
f(\l_{SYM}) = \frac{\sqrt{\l_{SYM}}}{\pi} ~+ {\cal O}\left(\lambda^0\right)
\eeq
whereas for the ABJM theory \cite{GV}-\cite{Astolfi}
\bea
\label{scalingCS}
f_{CS} (\lambda) = \sqrt{2\, \lambda} + {\cal O} \left(\lambda^0 \right)
\eea
it is easy to see that
\beq
\frac{R_{AdS_5}^2}{R_{AdS_4}^2} = \frac{f(\l_{SYM})}{f_{CS}(\l)}\Big|_{\rm leading}
\eeq

Moreover,  the string solution in $AdS_5$ describing a four--point amplitude/light--like WL for ${\cal N}=4$ SYM at strong coupling \cite{AM} may be straightforwardly embedded in $AdS_4$ as well.
Therefore the four--point amplitude should be trivially readable from the ${\cal N}=4$ SYM result (\ref{stramp}) by changing the $AdS$ radius.
Indeed this supports the extension of the BDS--like ansatz (\ref{3dBDS}) to strong coupling.

The subtle point in identifying the ${\cal N}=4$ solution with the ABJM one comes with regularization of infrared divergences. In \cite{AM}, the strong coupling analogue of dimensional regularization is spelled out. This amounts to continuing the dimensions of the $D_p$--branes sourcing the $AdS_5\times S^5$ background from $p=3$ to $p=(3-2\e)$.
Correspondingly, the new solution for the modified metric leads to  the expression
\bea\label{modified}
S_\e  = \frac{ \sqrt{\lambda_D c_{\epsilon} }}{ 2 \pi }\,  \int \frac{ {\cal L}_{\epsilon=0}}{ r^{\epsilon} }
\eea
for the regularized world--sheet action that, once minimized, will provide the four--point amplitude (here $c_{\epsilon}$ is an $\epsilon$ dependent constant and $\lambda_D$ is the dimensionless 't Hooft coupling in dimensional regularization).

In the context of ABJM we have not been able to find a similarly well--motivated regularization procedure \footnote{Although cutoff regularization works fine in three dimensions, we prefer to insist on a dimensional--like one in order to compare with our expression (\ref{result}).}.
However, guided by the analogy between the four--point ABJM and ${\cal N}=4$ SYM amplitudes at weak coupling, we are tempted to employ the prescription (\ref{modified}) to regularize the action also in the ${\rm AdS}_4$ context. Following the same steps as for the ${\cal N}=4$ SYM case, it leads to a strong coupling version of the ABJM four--point amplitude given by
\bea\label{strong}
{\cal M}_4   = e^{Div + \frac{\sqrt{2\lambda}}{8}\left(\ln^2\left(\frac{s}{t}\right)+ \frac{4 \pi^2}{3}\right) + {\rm Consts} + {\cal O}\left(\frac{1}{\sqrt{\lambda}}\right)}
\eea	
where the leading infrared divergence is
\bea
Div\big|_{leading} = - \frac{\sqrt{2}}{\e^2}\, \sqrt{\frac{\l\, \mu^{2\e}}{s^{\e}}} - \frac{\sqrt{2}}{\e^2}\, \sqrt{\frac{\l\, \mu^{2\e}}{t^{\e}}}
\eea
Even though this prescription lacks strong motivations, it definitely captures the essential features of the amplitude, such as the leading singularity and the coefficient of the non--trivial finite piece, which matches the strong coupling value of the ABJM scaling function (\ref{scalingCS}).

The generalization to the ABJ model is not straightforward. Here the situation is slightly subtler, since concerns have arisen on the integrability of the corresponding $\sigma$--model in the dual description \cite{MZ}.
Nevertheless, to first order at strong coupling we still expect the amplitude to be described by (\ref{strong}).
The reason is that unitarity requires $l=|M-N| < K$ \cite{ABJ}. Hence at strong coupling, where $M,N \gg K$, the shift in the ranks is negligible compared to $\sqrt{\l}$ of ABJM, and it affects the solution at higher orders only, starting from ${\cal O}\left(\frac{1}{\sqrt{\lambda}}\right)$ \cite{BH}.

\vskip 25pt
\section*{Acknowledgements}
\noindent

We thank  Wei-Ming Chen, Johannes Henn, Yu-tin Huang, Jan Plefka and
Konstantin Wiegandt for valuable discussions. This work has been supported in part by INFN and MIUR.

\vfill
\newpage
\appendix
\section{Notations and conventions}

We work in three dimensional euclidean ${\cal N}=2$ superspace described by coordinates $(x^\mu, \th^\alpha\, \thb^\beta)$, $\a, \b =1,2$ .

Spinorial indices are raised and lowered as (we follow conventions of \cite{superspace})
\begin{eqnarray}
  \psi^\alpha=C^{\alpha\beta}\psi_\beta  \qquad \psi_\alpha=\psi^\beta C_{\beta\alpha}
\end{eqnarray}
where the $C$ matrix
\begin{eqnarray}
  C^{\alpha\beta} = \left(\begin{array}{cc} 0 & i \\ -i & 0 \end{array}\right) \qquad
  C_{\alpha\beta} = \left(\begin{array}{cc} 0 & -i \\ i & 0 \end{array}\right)
\end{eqnarray}
obeys the relation
\begin{eqnarray}
  C^{\alpha\beta}\, C_{\gamma\delta}
  &=\delta^\alpha{}_\gamma\, \delta^\beta{}_\delta - \delta^\alpha{}_\delta\, \delta^\beta{}_\gamma
\end{eqnarray}
Spinors are contracted according to
\begin{eqnarray}
  \psi\chi=\psi^\alpha\, \chi_\alpha=\chi^\alpha\, \psi_\alpha=\chi\psi
  \qquad
  \psi^2=\frac{1}{2}\, \psi^\alpha\, \psi_\alpha
\end{eqnarray}

Dirac $\left(\g^{\mu}\right)^\alpha{}_\beta$ matrices are defined to satisfy the algebra
\begin{eqnarray}
  (\gamma^\mu)^\alpha{}_\gamma\, (\gamma^\nu)^\gamma{}_\beta
  =-g^{\mu\nu}\delta^\alpha{}_\beta + i\, \epsilon^{\mu\nu\rho}\, (\gamma_\rho)^\alpha{}_\beta
\end{eqnarray}
Trace identities needed for loop calculations can be easily obtained from the above algebra
\begin{eqnarray}
  &\tr(\gamma^\mu\, \gamma^\nu)
  &= (\gamma^\mu)^\alpha{}_\beta\, (\gamma^\nu)^\beta{}_\alpha
  = - 2\,  g^{\mu\nu} \\
  &\tr(\gamma^\mu\, \gamma^\nu\, \gamma^\rho)
  &= - (\gamma^\mu)^\alpha{}_\beta\, (\gamma^\nu)^\beta{}_\gamma\, (\gamma^\rho)^\gamma{}_\alpha
  = 2\,i\,\epsilon^{\mu\nu\rho} \\
  &\tr(\gamma^\mu\, \gamma^\nu\, \gamma^\rho\, \gamma^\sigma) &= (\gamma^\mu)^\alpha{}_\beta\, (\gamma^\nu)^\beta{}_\gamma\,
  (\gamma^\rho)^\gamma{}_\delta\, (\gamma^\sigma)^\delta{}_\alpha = \non\\
  &&= 2\, (g^{\mu\nu}\, g^{\rho\sigma}-g^{\mu\rho}\, g^{\nu\sigma}+g^{\mu\sigma}\, g^{\nu\rho})
\end{eqnarray}

Using these matrices, vectors and bispinors are exchanged according to
\begin{eqnarray}
\begin{array}{lll}
\mathrm{coordinates:}\qquad   &  x^\mu=(\gamma^\mu)_{\alpha\beta}\, x^{\alpha\beta}
    &\qquad
     x^{\alpha\beta} = \frac{1}{2}\, (\gamma_\mu)^{\alpha\beta}\, x^\mu \\
 \mathrm{derivatives:}\qquad    &   \partial_\mu=\frac{1}{2}\, (\gamma_\mu)^{\alpha\beta}\,
    \partial_{\alpha\beta} & \qquad
    \partial_{\alpha\beta}=(\gamma^\mu)_{\alpha\beta}\, \partial_\mu \\
\mathrm{fields:}\qquad   &
      A_\mu=\frac{1}{\sqrt{2}}\, (\gamma_\mu)^{\alpha\beta}\, A_{\alpha\beta}
      & \qquad A_{\alpha\beta}=\frac{1}{\sqrt{2}}\, (\gamma^\mu)_{\alpha\beta}\, A_\mu
\end{array}
\end{eqnarray}
It follows that the scalar product of two vectors can be rewritten as
 \begin{eqnarray}
p\cdot k \, = \, \frac12 \,   p^{\alpha\beta}\,k_{\alpha\beta}
\end{eqnarray}

Superspace covariant derivatives are defined as
\begin{equation}
  D_\a = \pa_\a + \frac{i}{2}\,  \thb^\b\,  \pa_{\a\b}
  \qquad , \qquad \Db_\a = \bar\pa_\a
  + \frac{i}{2}\,  \th^\b\,  \pa_{\a\b}
\end{equation}
and satisfy the anticommutator
\begin{equation}
\label{algebra}
  \{D_\a ,\,  \Db_\b\} = i\,  \pa_{\a\b}
\end{equation}

\bigskip
\noindent

The components of a chiral and an anti--chiral superfield,
$Z(x_L,\theta)$ and $\bar{Z}(x_R,\bar{\theta})$, are a complex boson
$\phi$, a complex two--component fermion $\psi$ and a complex auxiliary
scalar $F$. Their expansions are given by
\begin{eqnarray}
  Z = \phi(x_L) + \th^\a \psi_\a(x_L) - \th^2 \,
  F(x_L) \non \\ \bar{Z} = \bar{\phi}(x_R)
  + \thb^\a \bar{\psi}_\a(x_R) - \thb^2 \, \bar{F}(x_R)
\end{eqnarray}
where $x_L^\mu = x^\mu + i \theta \gamma^\mu \bar{\theta}$, $x_R^\mu =
x^\mu - i \theta \gamma^\mu \bar{\theta}$.

The components of the real vector superfield
$V(x,\theta,\bar{\theta})$ in the Wess-Zumino gauge ($V| = D_\a V| = D^2
V| = 0$) are the gauge field $A_{\a\b}$, a complex two--component
fermion $\lambda_\a$, a real scalar $\sigma$ and an auxiliary scalar
$D$, such that
\begin{equation}
  \label{eqn:WZgauge}
  V = i \, \th^\a \thb_\a \, \sigma(x)
  + \th^\a \thb^\b \, \sqrt{2} \, A_{\a\b}(x)
  - \th^2 \, \thb^\a \bar{\lambda}_\a(x)
  - \thb^2 \, \th^\a \lambda_\a(x)
  + \th^2 \, \thb^2 \, D(x)
\end{equation}

\bigskip
\noindent
The vector superfields $(V,\hat{V})$ are in the adjoint representation of the two gauge groups
$U(M) \times U(N)$, that is  $V = V_A T^A$ and $\hat{V} = \hat{V}_A \hat{T}^A$, where $T^A$ are
the $U(M)$ generators and $ \hat{T}^A$ are the $U(N)$ ones.

The $U(M)$ generators are defined as $T^A = (T^0, T^a)$, where $T^0 =
\frac{1}{\sqrt{N}}$ and $T^a$ ($a=1,\ldots, M^2-1$) are a set of
$M\times M$ hermitian matrices.  The generators are normalized as
$\Tr( T^A T^B )= \delta^{AB}$. The same conventions hold for the $U(N)$
generators.

\bigskip
\noindent
For any value of the couplings, the action (\ref{eqn:action}) is
invariant under the following gauge transformations
\begin{eqnarray}
  && e^V \rightarrow e^{i \bar{\L}_1} e^V e^{-i\L_1}
  \qquad \qquad e^{\hat{V}} \rightarrow e^{i\bar{\L}_2} e^{\hat{V}} e^{-i\L_2}
  \\
  && \non \\
  && A^i \rightarrow e^{i\L_1} A^i e^{-i\L_2}
  \qquad \qquad B_i \rightarrow e^{i\L_2} B_i e^{-i\L_1}
  \label{gaugetransf}
\end{eqnarray}
where $\L_1, \L_2$ are two chiral superfields parametrizing $U(M)$ and
$U(N)$ gauge transformations, respectively. Antichiral superfields
transform according to the conjugate of (\ref{gaugetransf}).  The
action is also invariant under the $U(1)_R$ R--symmetry group under
which the $A^i$ and $B_i$ fields have $\frac12$--charge.

\section{Integrals in dimensional regularization}

In this Appendix we list a number of properties for momentum integrals  entering the evaluation of four--point scattering amplitudes.

We work in dimensional regularization, $d=3-2\e$, with dimensional reduction (spinors and $\e$--tensors are kept strictly in three dimensions).

\subsection{Tools}

When evaluating one--loop bubbles and two--loop factorized and composite bubbles we used the $G[a,b]$ functions defined by
\beq
\label{G}
G[a,b] = \frac{1}{(4\pi)^{d/2}} \; \frac{\G(a+b-d/2) \, \G(d/2 - a) \, \G(d/2 - b)}{\G(a) \, \G(b) \, \G(d-a-b)}.
\eeq
Moreover, we introduce the compact notation $\Gamma(a_1|...|a_n)=\Gamma(a_1)...\Gamma(a_n)$.

In order to write a momentum integral in its Feynman parametrized form, the basic identity is
\begin{equation}
\frac{1}{A_1^{\alpha_1}}\dots\frac{1}{A_n^{\alpha_n}}=
\frac{\Gamma(\alpha_1+\dots+\alpha_n)}{\Gamma(\alpha_1\,|\,\dots\,|\,\alpha_n)}\int\limits_{0}^{1}
\frac{d\beta_1\dots d\beta_n\,\delta(\beta_1+\dots+\beta_n-1)\,\beta_1^{\alpha_1-1}\dots\beta_n^{\alpha_n-1}}
{(\beta_1\,A_1+\dots+\beta_n\,A_n)^{\alpha_1+\dots+\alpha_n}},
\end{equation}
where $A_j$ are generic propagators. The integration over Feynman parameters makes often use of the identity
\begin{equation}
\int\limits_{0}^{1}
d\beta_1\dots d\beta_n\,\delta(\beta_1+\dots+\beta_n-1)\,\beta_1^{\alpha_1-1}\dots\beta_n^{\alpha_n-1}=
\frac{\Gamma(\alpha_1\,|\,\dots\,|\,\alpha_n)}{\Gamma(\alpha_1+\dots+\alpha_n)}
\end{equation}

The most complicated computations of the one and two--loop on--shell amplitude were performed using Mellin-Barnes representations \cite{Davydychev,Smirnov}. These representations are based on the identity
\begin{equation}\label{0F1}
\frac{1}{(k^2+M^2)^a}=\frac{1}{(M^2)^a\Gamma(a)}\,\frac{1}{2\pi i}
\int\limits_{-i\infty}^{i\infty} d{\bf s}\,\Gamma(-{\bf s})\Gamma({\bf s}+a)\left(\frac{k^2}{M^2}\right)^{\bf s},
\end{equation}
where the contour is given by a straight line along the imaginary
axis such that indentations are used if necessary in order to leave
the series of poles ${\bf s}=0,1,\cdots,n$ to the right of the contour and
the series ${\bf s}=-a,-a-1,\cdots,-a-n$ to the left of the contour.

After Feynman--parametrizing a triangle integral and using (\ref{0F1}),
the following formula holds
\begin{align}
&\int\frac{d^d k}{(2\pi)^d}\frac{1}{k^{2\mu_1}(k-p)^{2\mu_2}(k+q)^{2\mu_3}}= \frac{1}{(4\pi)^{d/2}\prod_i\Gamma(\mu_i)\Gamma(d-\sum_i\mu_i)}\times\nonumber\\
&\times\int\limits_{-i\infty}^{i\infty}\frac{d{\bf s}\,d{\bf t}}{(2\pi i)^2}\,
\frac{\Gamma\left(-{\bf s}|\!-\!{\bf t}|\tfrac{d}{2}-\!\mu_1\!-\!\mu_2\!-\!{\bf s}|\tfrac{d}{2}-\!\mu_1\!-\!\mu_3\!-\!{\bf t}
|\mu_1\!+\!{\bf s}\!+\!{\bf t}|{\sum}_i\mu_i\!-\!\tfrac{d}{2}\!+\!{\bf s}+\!{\bf t}\right)}
{(p^2)^{-{\bf s}}\,(q^2)^{-{\bf t}}\,(p+q)^{2({\bf s}+{\bf t}+\sum_i\mu_i-d/2)}}
\end{align}
while for a vector--like triangle we have
\begin{align}
\label{MBvectortriangle}
&\int\frac{d^d k}{(2\pi)^d}\frac{k^{\nu}}{k^{2\mu_1}(k-p)^{2\mu_2}(k+q)^{2\mu_3}}=\nonumber\\
&=\frac{(4\pi)^{-d/2}}{\prod_i\Gamma(\mu_i)\Gamma(d-\sum_i\mu_i+1)}
\int\limits_{-i\infty}^{i\infty}\frac{d{\bf s}\,d{\bf t}}{(2\pi i)^2}\,
\frac{\Gamma\left(\!-{\bf s}|\!-\!{\bf t}|\mu_1\!+{\bf s}\!+{\bf t}|
{\sum}_i\mu_i\!-\!\tfrac{d}{2}\!+\!{\bf s}\!+\!{\bf t}\right)}
{(p^2)^{-{\bf s}}\,(q^2)^{-{\bf t}}\,(p+q)^{2({\bf s}+{\bf t}+\sum_i\mu_i-d/2)}}\times\nonumber\\
&\left[\Gamma(\tfrac{d}{2}\!-\!\mu_1\!-\!\mu_2\!-\!{\bf s}|\tfrac{d}{2}\!-\!\mu_1\!-\!\mu_3\!-\!{\bf t}\!+\!1)p^{\,\nu}\! -\!\Gamma(\tfrac{d}{2}\!-\!\mu_1\!-\!\mu_2\!-\!{\bf s}\!+\!1|\!\tfrac{d}{2}\!-\!\mu_1\!-\!\mu_3\!-\!{\bf t})q^{\,\nu}\right]
\end{align}
where the multiple contours are taken using the convention already mentioned for the relative position of the poles. When the position of a pole is chosen differently compared to the convention, it is customary to use the notation $\Gamma^{\star} (z)$ for the gamma function involved.

We can proceed along similar lines for writing the Mellin--Barnes representation for box diagrams, vector--like boxes, etc. Using these representations, along with the Barnes first lemma
\begin{equation}
\label{lemma}
\int\limits_{-i\infty}^{i \infty} \frac{d{\bf s}}{2\pi i}\,
\Gamma(a+{\bf s})\Gamma(b+{\bf s})\Gamma(c-{\bf s})\Gamma(d-{\bf s})=
\frac{\Gamma(a+c)\Gamma(a+d)\Gamma(b+c)\Gamma(b+d)}{\Gamma(a+b+c+d)}
\end{equation}
we have been able to compute one and two--loop amplitudes in a manifestly analytical way, without performing numerical evaluations.

\subsection{Bubbles}

At one--loop, the evaluation of simple bubbles is required.
Feynman parametrizing the integrand and working off--shell ($p^2 \neq 0$), we easily obtain
\beq
\label{bubble}
{\cal B}(p) \equiv \intke{k} \frac{1}{k^2 (k+p)^2} = G[1,1] \, \frac{1}{|p|^{1+2\e}}
\sim \frac{1}{8\,|p|} + {\cal O}(\e)
\eeq
where we have defined $|p| \equiv \sqrt{p^2}$ and $G[a,b]$ is given in (\ref{G}).
On the other hand, if we are on--shell, $p^2=0$, the integral reduces to a tadpole--like integral and in dimensional regularization it vanishes.

\subsection{Triangles}

\begin{figure}[h!]
    \centering
    \includegraphics[width=0.35\textwidth]{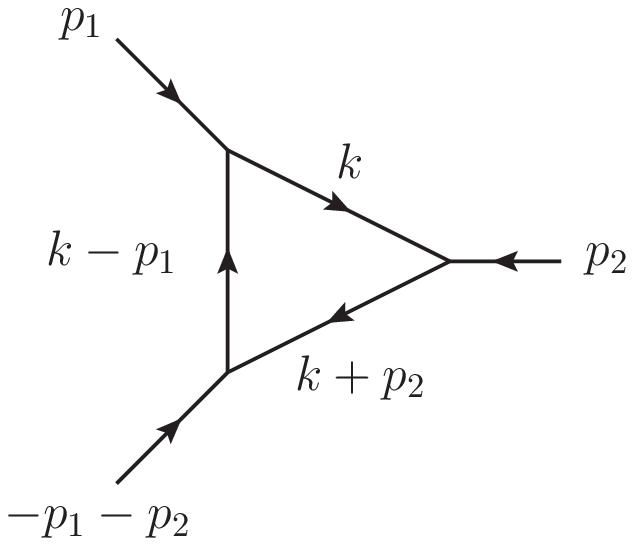}
    \caption{The triangle diagram.}
    \label{triangulos_scalar}
\end{figure}

We begin by evaluating the scalar triangle diagram of Fig. \ref{triangulos_scalar}.

When the external momenta are off-shell ($p_i^2 \neq 0$), the integral can be computed in three dimensions with no need for regularization. Since for $D=3$ the triangle with propagator exponents $(1,1,1)$ satisfies the uniqueness condition \cite{Usyukina:1994iw}, it evaluates to a rational function
\bea
\label{trioffshell}
\mathcal{T}(p_1,p_2) &\equiv& \int\frac{d^{3}k}{(2\pi)^{3}}\frac{1}{k^2\,(k-p_1)^2\,(k+p_2)^2}
\non \\
&=&  \frac{1}{8|p_1||p_2||p_1 + p_2|}
\eea
The corresponding integral, when evaluated on--shell ($p_i^2 = 0$) and in dimensional regularization, can be easily treated by Feynman parametrization and is given by
\bea
\label{triangle}
\mathcal{T}(p_1,p_2) &\equiv& \left.\int\frac{d^{3-2\epsilon}k}{(2\pi)^{3-2\epsilon}}\frac{1}{k^2\,(k-p_1)^2\,(k+p_2)^2}\right|_ {p_i^2=0}
\non \\
&=&  \frac{\Gamma(3/2+\epsilon)\Gamma^2(-1/2-\epsilon)}{(4\pi)^{\frac32-\e}\Gamma(-2\epsilon)}\frac{1}{|p_1 + p_2|^{3+2\epsilon}}
\non \\
& \sim  &  - \e \; \frac{1}{ 2|p_1 + p_2|^3} + {\cal O}(\e^2)
\eea
Therefore, in dimensional regularization and on--shell limit, the scalar triangle can be set to zero.

We then consider the on--shell, vector--like triangle. Again, by Feynman parametrization,
it is straightforward to show that the integral is given by
\bea
\label{vector-triangle}
\mathcal{T}_{\cal V}^{\a \b}(p_1,p_2) && \equiv \quad \left.\int\frac{d^{3-2\epsilon}k}{(2\pi)^{3-2\epsilon}}\frac{k^{\a\b}}{k^2\,(k-p_1)^2\,(k+p_2)^2}\right|_ {p_i^2=0}
\non \\
&& = \quad \frac{\Gamma(3/2+\epsilon)\Gamma(1/2-\epsilon)\Gamma(-1/2-\epsilon)}{(4\pi)^{\frac32 -\e}\Gamma(1-2\epsilon)} \ \frac{\left(p_1-p_2\right)^{\a\b}}{|p_1 + p_2|^{3+2\epsilon}}
\non \\
&& \xrightarrow{\e \rightarrow 0} \;  {\mathcal B} (p_1 + p_2) \; \frac{(p_2 - p_1)^{\a\b}}{(p_1 + p_2)^2}
\eea
where ${\mathcal B} (p_1 + p_2)$ is the bubble in eq. (\ref{bubble}).

An important observation is that, as a consequence of the on--shell conditions, the vector--like triangle satisfies the following identities
\begin{equation}
\label{triangle-id}
\mathcal{T}_{\cal V}(p_1,p_2) \cdot (p_1+p_2) = - \mathcal{T}_{\cal V}(p_1,p_2) \cdot (p_3+p_4)= 0
\end{equation}

\subsection{Boxes}

\begin{figure}[h!]
    \centering
    \includegraphics[width=0.38\textwidth]{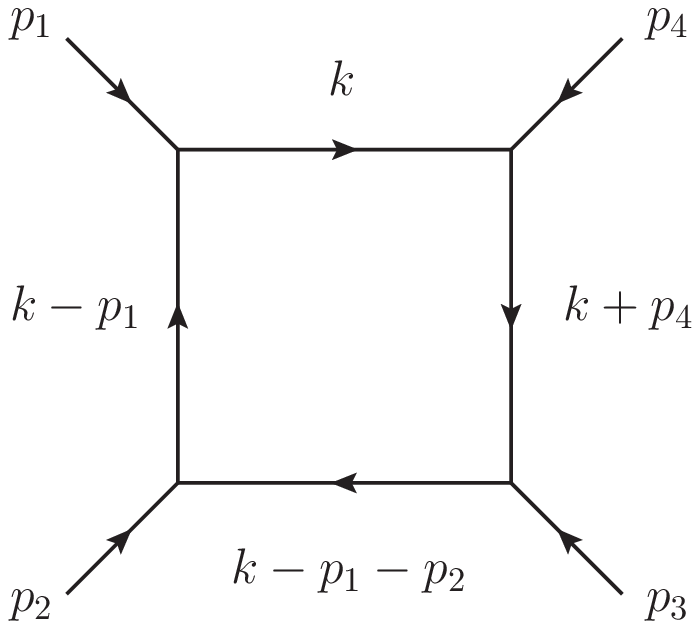}
    \caption{The box diagram.}
    \label{cuadrado}
\end{figure}

We now consider scalar and vector--like box diagrams drawn in Fig. \ref{cuadrado}.

\noindent
In terms of the Mandelstam variables, the scalar integral is written as
\begin{equation}
\label{scalarbox}
\mathcal{Q}(s,t)=\left.\int\frac{d^{3-2\epsilon}k}{(2\pi)^{3-2\epsilon}}\frac{1}{k^2\,(k-p_1)^2\,(k-p_1-p_2)^2
\,(k+p_4)^2}\right|_ {p_i^2=0}
\end{equation}
while the vector--like integral is
\beq
\label{vectorbox}
\mathcal{Q}_{\cal V}^{\mu} =\left.\int\frac{d^{3-2\epsilon}k}{(2\pi)^{3-2\epsilon}}\frac{k^\mu}{k^2\,(k-p_1)^2\,(k-p_1-p_2)^2\,(k+p_4)^2}\right|_ {p_i^2=0}
\end{equation}

The scalar integral can be evaluated at leading order in $\epsilon$. Feynman parametrizing the integrand in eq. (\ref{scalarbox})  we obtain
\begin{equation}
\mathcal{Q}(s,t) = \frac{\Gamma(5/2+\epsilon)}{(4\pi)^{3/2-\epsilon}}\int\frac{dy_1\,dy_2\,dy_3\,dy_4\,\delta(\sum_i y_i-1)}{(y_1\,y_3\, s+y_2\,y_4\, t)^{5/2+\epsilon}}
\end{equation}
Expressing the denominator as a Mellin Barnes integral
\begin{equation}
\frac{1}{(y_1\,y_3\, s+y_2\,y_4\,t)^{5/2+\epsilon}}=\frac{1}{\Gamma(5/2+\epsilon)}
\int \frac{du}{2\pi i}\Gamma(-u)\Gamma(u+5/2+\epsilon)\frac{(y_1\,y_3\,s)^u}{(y_2\,y_4\,t)^{5/2+\epsilon+u}}
\end{equation}
and integrating on the Feynman parameters we obtain a one--fold representation
\begin{equation}
\frac{2\, \epsilon \, (1+2\epsilon)}{(4\pi)^{D/2}\, \Gamma(1-2\epsilon)\, t^{5/2+\epsilon}}\int\frac{du}{2\pi i}
\Gamma(-u)\Gamma^2(-3/2-\epsilon-u)\Gamma(5/2+\epsilon+u)\Gamma^2(1+u){X}^u
\end{equation}
We note that the MB integral is multiplied by an $\epsilon$ factor and the integral itself is well defined when $\epsilon\to 0$. Therefore, to leading order in $\epsilon$ we have
\begin{align}
\mathcal{Q}(s,t)&=
\frac{\epsilon}{4\pi^{3/2}\,t^{5/2}}\int\frac{du}{2\pi i}
\Gamma(-u)\Gamma^2(-3/2-u)\Gamma(5/2+u)\Gamma^2(1+u){X}^u+\mathcal{O}(\epsilon^2)
\nonumber\\
&\equiv \frac{\epsilon}{4\pi^{3/2}\,t^{5/2}}(f_1(X)+f_2(X))+\mathcal{O}(\epsilon^2)
\end{align}
where $X=s/t$. By closing the contour on the right,  $f_1(X)$ is the sum of the residues at the poles of $\Gamma(-u)$, whereas $f_2(X)$ is the contribution from the poles of  $\Gamma^2(-3/2-u)$.

The $f_1(X)$ function is easily computed and gives
 \begin{equation}
f_1(X)=\pi^2\sum\limits_{n=0}^{\infty}\frac{(-X)^n\,n!}{\Gamma(5/2+n)}=
4\pi^{3/2}\left(\frac{\sqrt{1+X}}{X^{3/2}}\ln(\sqrt{X}+\sqrt{1+X})-\frac{1}{X}\right)
\end{equation}
The $f_2(X)$ function is more complicated since we have to deal with double poles. A first set of  manipulations leads to
\begin{equation}\label{f2}
f_2(X)=\frac{\pi}{X^{3/2}}\sum\limits_{n=0}^{\infty}\frac{\Gamma(-1/2+n)(-X)^n}{n!}\left(\ln X + \Psi(-1/2+n)-\Psi(1+n)\right)
\end{equation}
where $\Psi(x)$ is the digamma function.

The first term in (\ref{f2}) is easily summed to
\begin{equation}
\label{partial1}
\sum\limits_{n=0}^{\infty}\frac{(-X)^n\,\Gamma(-1/2+n)}{n!}=-2\sqrt{\pi}\,\sqrt{1+X}
\end{equation}

The second series in (\ref{f2}) can be summed by using the trick
\begin{align}
\label{partial2}
&\sum\limits_{n=0}^{\infty}\frac{(-X)^n\,\Gamma(-1/2+n)\Psi(-1/2+n)}{n!}=\frac{d}{d a}
\left(\sum\limits_{n=0}^{\infty}\frac{(-X)^n\,\Gamma(a+n)}{n!}\right)_{a=-1/2}\nonumber\\
&=\frac{d}{da}\left(\frac{\Gamma(a)}{(1+X)^a}\right)_{a=-1/2}=2\sqrt{\pi}\sqrt{1+X}
\left(\ln(1+X)-\Psi(-1/2)\right)
\end{align}

For the third term in (\ref{f2}) we use the following identity
\begin{equation}
\Psi(1+n)=-\gamma_E+\int\limits_{0}^1 \frac{1-t^n}{1-t}\,dt
\end{equation}
to rewrite the digamma function inside the series. By exchanging the order of the sum and the integral and summing the series, we obtain
\bea
\label{partial3}
&& \sum\limits_{n=0}^{\infty}\frac{(-X)^n\,\Gamma(-1/2+n)\Psi(1+n)}{n!}
=2\sqrt{\pi}\left(
\gamma_E\,\sqrt{1+X}-\int\limits_0^1 dt\,\frac{\sqrt{1+X}-\sqrt{1+tX}}{1-t}\right)
\nonumber \\
&& \qquad  \qquad = -2\sqrt{\pi}\sqrt{1+X}\left(\Psi(-1/2)+2\ln(1+\tfrac{1}{\sqrt{1+X}})\right)+4\sqrt{\pi}
\eea
where the integral in the first line has been performed using \textit{Mathematica}.

Summing (\ref{partial1}, \ref{partial2}, \ref{partial3}), after many non--trivial cancelations and simplifications we obtain as a final result
\begin{equation}
\mathcal{Q}(s,t)=\frac{\epsilon}{(\sqrt{s\,t})^3}
\left[\sqrt{s+t}\,\ln\left(\frac{\sqrt{s}+\sqrt{t}+\sqrt{s+t}}{\sqrt{s}+\sqrt{t}-\sqrt{s+t}}\right)-(\sqrt{s}+\sqrt{t})\right]
+\mathcal{O}(\epsilon^2)
\end{equation}

The vector--like box integral (\ref{vectorbox}) can be computed by using the same methods as before or, equivalently, by using Passarino--Veltman reduction to write it as a linear combination of scalar integrals. In any case, at leading order in $\e$, we obtain
\begin{align}
\label{vector-box}
\mathcal{Q}_{\cal V}^\mu=\frac{\epsilon}{2(\sqrt{s\,t})^3}
&\Bigg\{ \frac{1}{\sqrt{s+t}} \, \ln\left(\frac{\sqrt{s}+\sqrt{t}+\sqrt{s+t}}{\sqrt{s}+\sqrt{t}-\sqrt{s+t}}\right)
\left[s\,(p_1-p_4)^\mu+t\,(p_1+p_2)^\mu\right]
\nonumber\\
&\quad -\sqrt{t}\,(p_1-p_4)^\mu-\sqrt{s}\,(p_1+p_2)^\mu\Bigg\}
+\mathcal{O}(\epsilon^2)
\end{align}
It is interesting to note that the projections of $\mathcal{Q}_{\cal V}$ in the directions of $p_1$ and $p_4$  become very simple
\begin{equation}
\label{projections}
\mathcal{Q}.p_1=\frac{\epsilon}{4}\left(\frac{1}{s^{3/2}}-\frac{1}{t^{3/2}}\right)+\mathcal{O}(\epsilon^2)\quad\mbox{and}\quad
\mathcal{Q}.p_4=\frac{\epsilon}{4}\left(\frac{1}{t^{3/2}}-\frac{1}{s^{3/2}}\right)+\mathcal{O}(\epsilon^2)
\end{equation}
since the logarithm term drops.

In fact, these two projections, can be calculated to all orders in $\epsilon$. Writing $k \cdot p_1$ and $k \cdot p_4$ in the numerator of (\ref{vectorbox}) as  the difference of two squares, the integral reduces to the difference of two triangles. Therefore, using the results of Subsection B.2, we obtain
\begin{align}
&\mathcal{Q}.p_1=\frac{\Gamma(3/2+\epsilon)\Gamma^2(-1/2-\epsilon)}{2 \, (4\pi)^{D/2}\Gamma(-2\epsilon)}
\left(\frac{1}{t^{3/2+\epsilon}}-\frac{1}{s^{3/2+\epsilon}}\right)
\nonumber\\
&\mathcal{Q}.p_4=\frac{\Gamma(3/2+\epsilon)\Gamma^2(-1/2-\epsilon)}{2\, (4\pi)^{D/2}\Gamma(-2\epsilon)}
\left(\frac{1}{s^{3/2+\epsilon}}-\frac{1}{t^{3/2+\epsilon}}\right).
\end{align}
Since the leading term for $\e \to 0$ coincides with the expressions (\ref{projections}), this is a consistency check
of our results.

\section{The basis of dual conformally invariant integrals}

In this Appendix we give the proof that the linear combination $I_{5s}  \equiv I_{1s}-I_{4s}$ of scalar integrals defined in (\ref{dual1}, \ref{dual4}), when evaluated in three dimensions, is free from IR divergences. This allows to conclude that the actual basis for two--loop amplitudes is $I_{1s}, I_{2s}, I_{3s}, I_{5s}$ plus their $t$--counterparts.

We consider $I_{4s}$ in (\ref{dual4}) and apply the following identity
\beq
x_{46}^2 = x_{56}^2 + x_{45}^2 + 2\, x_{45} \cdot x_{56}
\eeq
to its numerator, thus decomposing it as (we neglect the factor $x_{13}^2$)
\begin{figure}[h!]
    \centering
    \includegraphics[width=\textwidth]{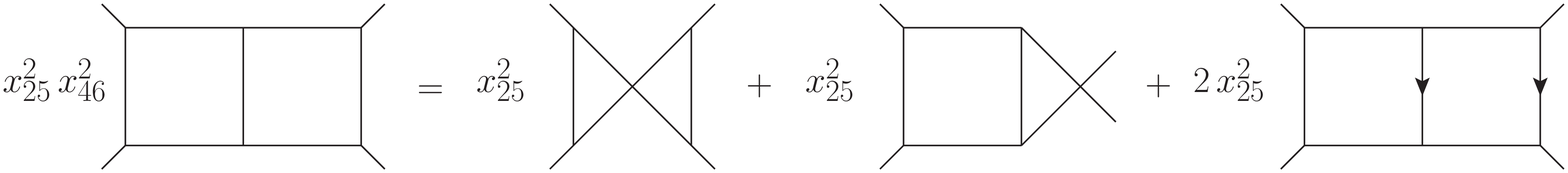}
\end{figure}

\noindent
where arrows indicate contractions of the corresponding variables at the numerator (see Fig. \ref{1set} for the labeling of momenta and dual variables).
Here we already recognize the emergence of the infrared divergent integral $I_{1s}$, a triangle--box which could diverge, having unprotected cubic vertices, and a double--box whose cubic vertices are mitigated by the presence of a non--trivial numerator.

We focus on the triangle--box and handle it by using the identity
\beq
x_{25}^2 = x_{56}^2 + x_{26}^2 - 2 x_{56} \cdot x_{26}
\eeq
in the numerator factor. The final result can be cast into the following graphical relation
\begin{figure}[h!]
    \centering
    \includegraphics[width=\textwidth]{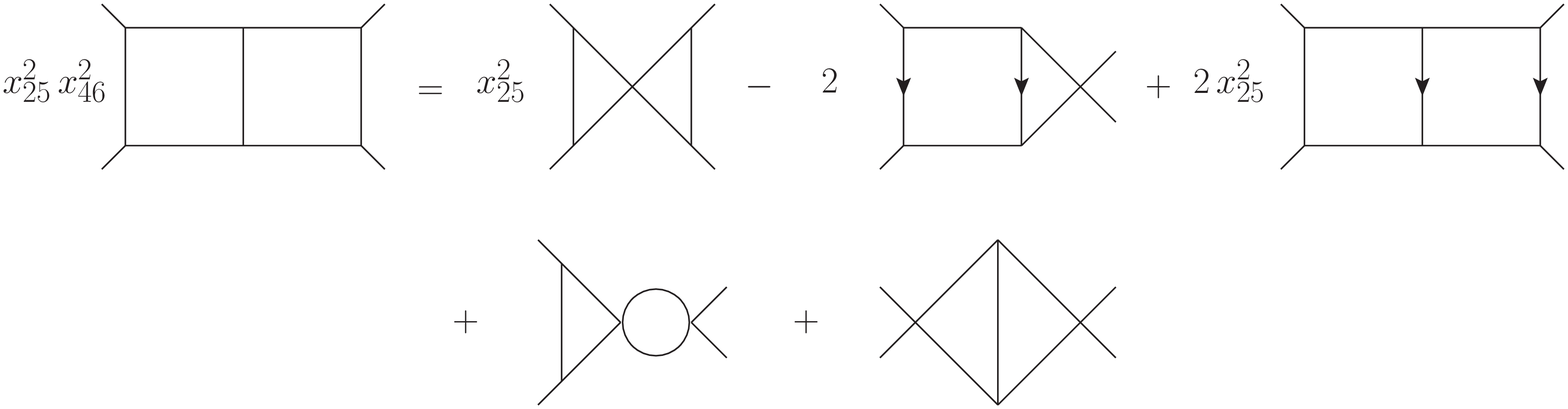}
       \label{id2}
\end{figure}

\noindent
where the IR divergence has been completely isolated in the last term. Therefore, taking the linear combination
$I_{1s} - I_{4s}$ the divergences cancel at the level of the integrands and we are left with a well--defined dual conformally invariant integral.

\end{document}